\documentclass[11pt,a4paper]{article}
\pdfoutput=1
\usepackage{jcappub}
\allowdisplaybreaks
\usepackage{footnote}	
\usepackage{xcolor}
\usepackage{bigints}
\usepackage{multirow}
\usepackage[normalem]{ulem}

\usepackage{graphicx,psfrag}
\usepackage{bm}
\usepackage{mathbbol,verbatim}
\usepackage{slashed}
\usepackage{graphics}

\graphicspath{{figs/}} 

\newcommand{\blue}{\textcolor{blue}}

\UseRawInputEncoding
\begin{document}

\title{Improved stellar limits on a light CP-even scalar}

\author[1,2]{Shyam Balaji,}
\author[3]{P. S. Bhupal Dev,}
\author[2, 4, 5]{Joseph Silk,}
\author[6]{and Yongchao Zhang}

\affiliation[1]{Laboratoire de Physique Th\'{e}orique et Hautes Energies (LPTHE), \\
	UMR 7589 CNRS \& Sorbonne Universit\'{e}, 4 Place Jussieu, F-75252, Paris, France}
	\affiliation[2]{Institut d¡¯Astrophysique de Paris, UMR 7095 CNRS \& Sorbonne Universit\'{e}, 98 bis boulevard Arago, F-75014 Paris, France}
\affiliation[3]{Department of Physics and McDonnell Center for the Space Sciences, Washington University, St. Louis, MO 63130, U.S.A}
\affiliation[4]{Department of Physics and Astronomy, The Johns Hopkins University, 3400 N. Charles	Street, Baltimore, MD 21218, U.S.A.}
\affiliation[5]{Beecroft Institute for Particle Astrophysics and Cosmology, University of Oxford, Keble	Road, Oxford OX1 3RH, U.K.}
\affiliation[6]{School of Physics, Southeast University, Nanjing 211189, China}

\emailAdd{sbalaji@lpthe.jussieu.fr}
\emailAdd{bdev@wustl.edu}
\emailAdd{silk@iap.fr}
\emailAdd{zhangyongchao@seu.edu.cn}

\abstract{We derive improved stellar luminosity limits on a generic light CP-even scalar field $S$ mixing with the Standard Model (SM) Higgs boson from the supernova SN1987A, the Sun, red giants (RGs) and white dwarfs (WDs). For the first time, we include the geometric effects for the decay and absorption of $S$ particles in the stellar interior. For SN1987A and the Sun, we also take into account the detailed stellar profiles. We find that a broad range of the scalar mass and mixing angle can be excluded by our updated astrophysical constraints. For instance, SN1987A excludes $1.0\times10^{-7} \lesssim \sin\theta \lesssim 3.0\times 10^{-5}$ and scalar mass up to 219 MeV, which covers the cosmological blind spot with a high reheating temperature. The updated solar limit excludes the mixing angle in the range of $1.5\times 10^{-12} < \sin\theta < 1$, with scalar mass up to 45 keV. The RG and WD limits are updated to $5.3\times 10^{-13} < \sin \theta < 0.39$ and $2.8\times 10^{-18} < \sin \theta < 1.8\times 10^{-4}$, with scalar mass up to 392 keV and 290 keV, respectively.}

\maketitle

\section{Introduction}

Possibilities for beyond the Standard Model (BSM) particles exhibit a wide range of freedom at the current juncture in high energy physics. They may be very heavy, for example at or above the TeV-scale, or they may conversely be very light, at or below the GeV-scale with very small or even feeble couplings. The latter scenario can be searched for in a large variety of terrestrial high-intensity experiments~\cite{Jaeckel:2010ni, Graham:2015ouw, Agrawal:2021dbo}. The observations of compact astrophysical objects can also provide complementary constraints on the couplings of light BSM particles, for instance from the luminosity observations of SN1987A, the Sun, red giants (RGs), horizontal-branch (HB) stars and white dwarfs (WDs)~\cite{Raffelt:1996wa}. In particular, if sufficiently light, BSM particles can be abundantly produced in the stars from interactions with nuclei, electrons, photons, etc. After being produced, these particles may decay into lighter standard model (SM) and/or other BSM particles, or get absorbed by the stellar medium. However, for certain ranges of couplings to the SM particles, these particles may leave the stellar material and take away appreciable energy from the cores, thus affecting stellar evolution or contradicting existing astrophysical observations. The simplest and most na\"{i}ve stellar limits on BSM particles are from luminosity observations, such as the solar luminosity~\cite{Song:2017kvf} and the observed neutrino luminosity from SN1987A~\cite{Hirata:1987hu}, using the so-called Raffelt criterion \cite{Raffelt:1996wa}.

In this paper, we focus on the stellar and supernova limits on a light CP-even scalar $S$, which is one of the most commonly studied BSM scenarios. The simplest underlying model is the singlet extension of the SM scalar sector, which can help stabilize the SM vacuum~\cite{Gonderinger:2009jp, Gonderinger:2012rd, Lebedev:2012zw, Elias-Miro:2012eoi, Khan:2014kba, Falkowski:2015iwa, Ghorbani:2021rgs}, address the hierarchy problem in relaxion models~\cite{Graham:2015cka, Flacke:2016szy, Frugiuele:2018coc, Banerjee:2020kww, Brax:2021rwk}, generate the baryon asymmetry in the Universe via baryogenesis~\cite{Espinosa:1993bs, Choi:1993cv, Ham:2004cf,Profumo:2007wc,Espinosa:2011ax, Barger:2011vm,  Profumo:2014opa,Curtin:2014jma, Kotwal:2016tex,Chen:2017qcz,Balaji:2020yrx}, address the cosmological constant problem through radiative breaking of classical scale invariance~\cite{Foot:2011et, Heikinheimo:2013fta, Wang:2015cda}, and mediate the interactions between dark matter (DM) and the SM sector~\cite{Pospelov:2007mp, Baek:2011aa, Baek:2012uj, Baek:2012se, Schmidt-Hoberg:2013hba, Krnjaic:2015mbs, Beniwal:2015sdl,Balaji:2018qyo}. With mass at the keV-scale, $S$ can also play the role of light DM~\cite{Babu:2014pxa,Balaji:2019fxd} and explain the anomalous X-ray spectrum at 3.55 keV~\cite{Bulbul:2014sua, Boyarsky:2014jta}. In this paper, we consider a generic scalar $S$, coupling to the SM particles through mixing with the SM Higgs $h$, parameterized by a single mixing angle $\sin\theta$.\footnote{It should be noted that for a leptonic or flavorful scalar, for instance coupling predominantly  to muons, the stellar limits may differ significantly~\cite{Bollig:2020xdr,Croon:2020lrf}. In some specific models, the mixing of CP-even scalars could also induce couplings of the associate Goldstone boson with the SM particles, which lead to some additional stellar limits~\cite{Keung:2013mfa, Tu:2015lwv}. }

In the supernova core, the production of $S$ is dominated by the nucleon bremsstrahlung process $N + N \to N + N + S$, with $N$ referring to both protons ($p$) and neutrons ($n$). The corresponding SN1987A limits on the light CP-even scalar $S$ have been studied in Refs.~\cite{Ishizuka:1989ts, Arndt:2002yg, Diener:2013xpa, Krnjaic:2015mbs, Tu:2017dhl, Lee:2018lcj, Dev:2020eam, Dev:2021kje}. Assuming constant baryon density $n_B = 1.2 \times 10^{38} \ {\rm cm}^{-3}$ and temperature  $T=30$ MeV within the supernova core with radius $R_c = 10$ km, and applying the observed neutrino luminosity of ${\cal L}_\nu = 3\times 10^{53}$ erg/sec, the mixing angle $\sin\theta$ is excluded in the range of $5.9 \times 10^{-7}$ to $7.0 \times 10^{-6}$ with scalar mass up to 148 MeV~\cite{Dev:2020eam}. In the Sun, RGs, HB stars and WDs, the temperatures are much lower, i.e. around the keV-scale. In this case, the light scalar $S$ will be produced predominantly from the $e-N_i$ bremsstrahlung process~\cite{Dev:2020jkh}, with $N_i$ labeling all the relevant nuclei.
Assuming constant baryon number density $n_B$, temperature $T$ and nuclei mass fraction $Y_{N_i}$,  the mixing angle in the range of $7.0 \times 10^{-18}$ to $1.2 \times 10^{-3}$ is excluded, with scalar mass up to 350 keV~\cite{Dev:2020jkh}.

In obtaining these luminosity limits,  constant $n_B$, $T$ and $Y_{N_i}$ were assumed, which is an over-simplification. In general, they are all functions of the radial distance $r$ from the surface to the center of the star, even if the star is assumed to be spherically symmetric. The phenomenological implications of stellar profiles have been investigated for stellar limits on dark photons~\cite{Rrapaj:2015wgs, Chang:2016ntp, Sung:2019xie, Sieverding:2021jfa, Hook:2021ous}, QCD axions or axion-like particles (ALPs)~\cite{Bollig:2020xdr,Croon:2020lrf, Lee:2018lcj, Hook:2021ous, Payez:2014xsa, Fischer:2016cyd, Chang:2018rso, Beznogov:2018fda, Dessert:2019sgw, Carenza:2019pxu, Bar:2019ifz, Lucente:2020whw, Fortin:2021cog, Caputo:2021kcv, Mori:2021krp, Fischer:2021jfm, Fiorillo:2021gsw}, $Z'$ bosons~\cite{Croon:2020lrf, Shin:2021bvz}, sterile neutrinos~\cite{Warren:2014qza, Syvolap:2019dat}, and dark or millicharged particles~\cite{Chang:2018rso, Guha:2018mli, Chu:2019rok, Sung:2021swd, Alonso-Alvarez:2021oaj}. Alternatively, shell models for the supernova core have been used to set limits on the $Z'$ boson and a light scalar coupling to both nucleons and neutrinos, where within each shell the density and temperature are constants~\cite{Cerdeno:2021cdz}. In  this paper, we include the stellar profiles for supernovae and the Sun, i.e. the density $\rho(r)$, nucleus mass fraction $Y_{N_i} (r)$ and temperature $T(r)$ as functions of $r$ 
in the context of the CP-even scalar $S$. For SN1987A, we adopt the numerical profiles Fischer $11.8M_\odot$, Fischer $18M_\odot$~\cite{Fischer:2016cyd} and Nakazato 13$M_\odot$ (with $M_\odot$ denoting the solar mass)~\cite{Nakazato:2012qf}, while for the Sun we consider the standard solar model~\cite{Bahcall:2004fg}. As a result, the mean free paths (MFPs) $\lambda(r)$ due to the absorption of $S$ in the stars are also functions of $r$. 
For the RGs and WDs, the stellar profiles suffer from much larger uncertainties than for SN1987A and the Sun, therefore we will use the constant densities and temperatures and recast the analysis done in Ref.~\cite{Dev:2020jkh}.\footnote{The limits from HB stars are expected to be similar to those from RGs, therefore we will not consider HB stars in this paper~\cite{Dev:2020jkh}. We have also not considered the impact of light scalar on the stellar evolution~\cite{Raffelt:1987yu, Ayala:2014pea} (e.g. ratio of the HB star to RG population in globular clusters), which could give additional constraints.}

After production of $S$,
the probability $P_{\rm decay}$ for $S$ to escape from the star before decaying into SM particles and the probability $P_{\rm abs}$ for $S$ to remain unabsorbed in the star depend on the geometry, namely where it is produced and its flight direction in the star. For instance, if $S$ is produced near the stellar surface and moves outward, it has a shorter distance to traverse before escaping from the star. On the other hand, if $S$ is produced near the surface but moves towards to the center of the star, the trajectory of $S$ tends to be much longer, and it is more easily decayed or absorbed in the star before it gets the chance to escape. Such a geometric factor has been (partially) taken into account in the supernova limits on axions~\cite{Burrows:1990pk,Lee:2018lcj}, dark photons~\cite{Chang:2016ntp, Rrapaj:2015wgs} and other dark particles~\cite{Sung:2021swd, Alonso-Alvarez:2021oaj} (see also Ref.~\cite{vanRiper:1981mko} for the leakage approximation). In this paper, we perform for the first time a careful treatment of the geometric factor for the decay and absorption of $S$ in supernovae, the Sun, RGs and WDs, which turns out to be important for the limits.

We then make a comparison of the updated stellar limits on the scalar mass $m_S$ and the mixing angle $\sin\theta$ obtained in this paper to the  na\"{i}ve results from Refs.~\cite{Dev:2020eam, Dev:2020jkh}. Here are  some of the important features that emerge:
\begin{itemize}
    \item For the three supernova profiles adopted in this paper, the resultant limits on $m_S$ and $\sin\theta$ are very similar, and exclude the range of mixing angle $1.0\times 10^{-7} \lesssim \sin\theta \lesssim 3\times10^{-5}$, which is significantly broader than that in the constant density and temperature case. The scalar mass up to 219 MeV is excluded, which is close to the constant profile case.
    \item With the standard solar model and the geometric factor included, the mixing angle in the range $1.5\times10^{-12}<\sin\theta <1$ is excluded, depending on the scalar mass up to 45 keV. In comparison with the constant profile case~\cite{Dev:2020jkh}, the limit on $\sin\theta$ is moved upward significantly.
    \item Including only the geometric factor, the RGs and WDs can exclude the mixing angle in the ranges of $5.3\times10^{-13} < \sin\theta < 0.39$ and $2.8\times10^{-18} < \sin\theta < 1.8\times 10^{-4}$, respectively, with scalar mass $m_S$ up to roughly 392 keV and 290 keV. Compared to the constant profile case~\cite{Dev:2020jkh}, a broader range of the mixing $\sin\theta$ is excluded, in particular the relatively large values.
\end{itemize}


The rest of the paper is organized as follows: Section~\ref{sec:supernova} focuses on the supernova limits. We first briefly introduce the supernova profiles in Section~\ref{sec:snprofiles}, and then describe the calculation details in Section~\ref{sec:sn:calculation}.  The updated supernova limits are obtained in Section~\ref{sec:sn:results}. Section~\ref{sec:Sun} mainly focuses on updating the solar limit on $S$. After sketching the standard solar model in Section~\ref{sec:sun:ssm}, the calculation details are given in Section~\ref{sec:sun:production}, and the resultant updated solar luminosity limit on $S$ is given in Section~\ref{sec:sn:results}. The updated stellar limits on $S$ from RGs and WDs can be found in Section~\ref{sec:others}. Finally, we summarize our results in Section~\ref{sec:conclusion}. Some important functions are collected in Appendix~\ref{sec:appendix}, and the emission rates for the subdominant production channels in the Sun are summarized in Appendix~\ref{sec:appendix2}.

\section{$S$ in the supernova}
\label{sec:supernova}

\subsection{Supernova profiles}
\label{sec:snprofiles}

There have been extensive studies of the astrophysical aspects of SN1987A.
It is an excellent candidate for examining new physics models because of the combination of the unique physical conditions attained in the star and the proximity of the explosion to our solar system. Despite this, constraints on new physics from the observation of SN1987A are inherently limited due to the difficulties associated with understanding the detailed physical processes of the supernova, even in the case without BSM physics. The main challenge in using SN1987A to constrain new physics is due to uncertainty surrounding the nature of the progenitor proto-neutron star which comprises the primary driver of the ``shock revival'' required to sustain the ultimate explosion. The mass of the progenitor star is uncertain up to a factor of two, and consequently the temperature and density profiles have large, qualitative uncertainties~\cite{Sukhbold:2015wba}.

For the purpose of illustrating the profile dependence of the supernova limits on $S$, we adopt the numerical profiles Fischer $11.8M_\odot$, Fischer $18M_\odot$~\cite{Fischer:2016cyd} and Nakazato 13$M_\odot$~\cite{Nakazato:2012qf}, which correspond respectively to  11.8, 18 and 13 solar mass progenitor stars at the time of 1 sec after collapse. The Nakazato 13$M_\odot$ profile is simulated with a 100 ms shock revival time inserted by hand. The Fischer 11.8$M_\odot$ and 18$M_\odot$ profiles are computed by solving the Boltzmann equation for neutrino transport with the {\tt AGILE-BOLTZTRAN} code~\cite{Liebendoerfer:2002xn} and an equation of state based on known nuclear isotopes and relativistic mean field models~\cite{Fischer:2016cyd}.

A summary of some of the benchmark physical parameters for the three SN1987A models of interest is shown in Table~\ref{tab:summary} (reproduced from Table~2 of Ref.~\cite{Chang:2016ntp}). These models can predict densities that vary by as much as an order of magnitude in certain regions of the proto-neutron star. These uncertainties simply imply that there is currently not full control of subtleties resulting from the behavior of the nuclear matter in this violent environment \cite{Chang:2016ntp}. Hence it follows that new physics constraints derived with any of the aforementioned profiles are necessarily approximate. However, the application of numerical SN1987A profiles represents significant improvements over modeling the supernova as a constant density and temperature object. Additional observations of core collapse could provide improved understanding of the nature of supernova cores in the future.

\begin{table}[!t]
    \centering
      \caption{Benchmark physical parameters for the three SN1987A models of interest. Here $R_\nu$ is the neutrinosphere radius, and $R_c$,  $\rho_c$, and $T_c$ refer to the core radius (of peak temperature), density and temperature  respectively~\cite{Chang:2016ntp}.}
    \label{tab:summary}
    \vspace{0.2cm}
    \begin{tabular}{l|c|c|c}
    \hline\hline
   Parameter & Fischer $11.8M_\odot$ & Fischer $18M_\odot$ & Nakazato $13M_\odot$ \\ \hline
    $R_\nu$ [km] & 24.9 & 23.6 & 25.6 \\
    $R_c$ [km] & $\sim 10$ & $\sim 11$ & $\sim 13$ \\
    $\rho_c$ [$10^{14}$ g/cm$^3$] & $\sim 1.8$ & $\sim 1.2$ & $\sim 1.0$ \\
    $T_c$ [MeV] & $\sim 29$ & $\sim 36$ & $\sim 34$ \\ \hline\hline
    \end{tabular}
    \end{table}

\begin{figure}[!t]
    \centering
    \includegraphics[height=0.24\textheight]{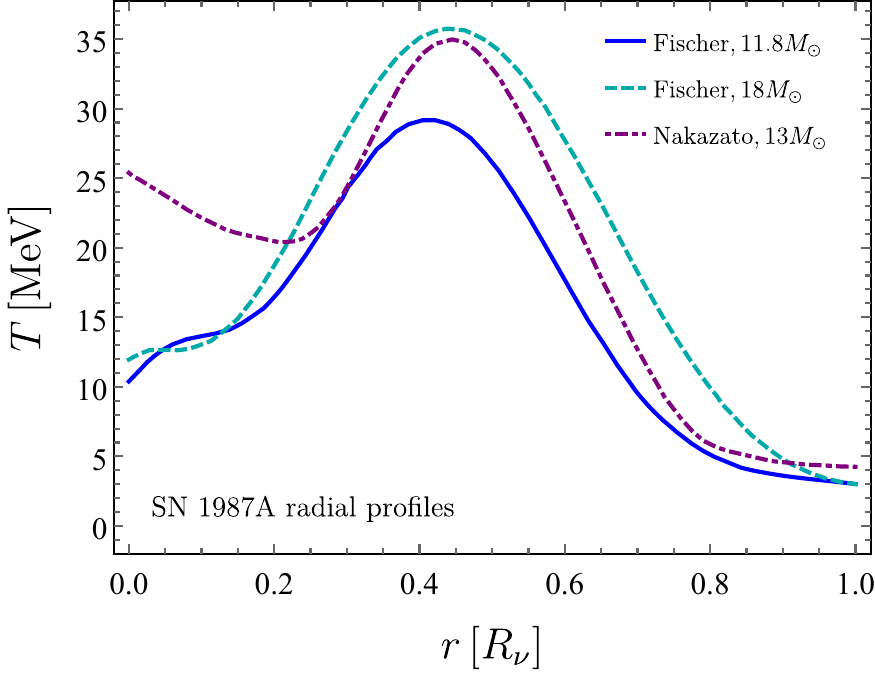}
    \includegraphics[height=0.24\textheight]{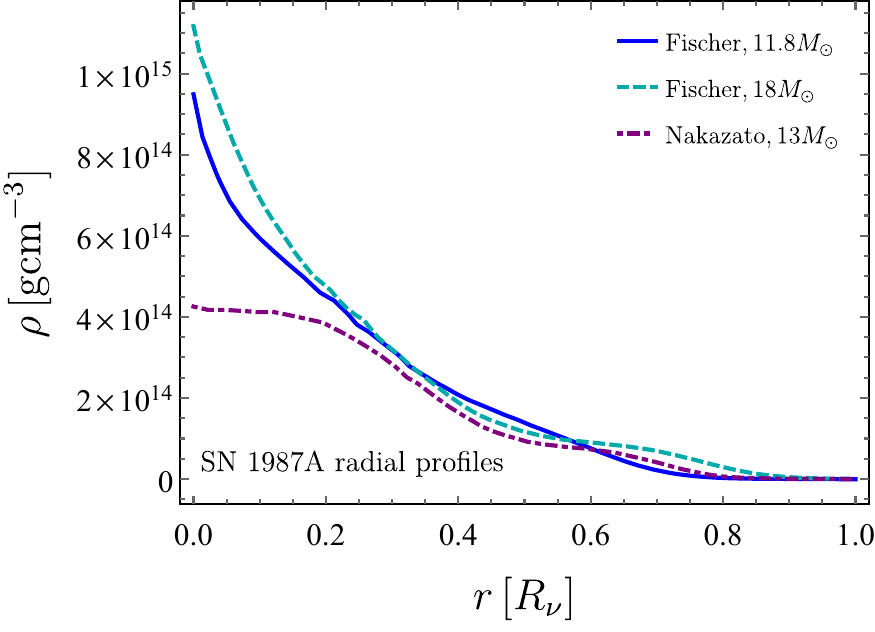}\\
    \caption{The profiles for temperature (left) and density (right) as functions of the radius $r$ for SN1987A. The Fischer $11.8M_\odot$, Fischer $18M_\odot$ and Nakazato $13M_\odot$ profiles are shown in both panels as the blue solid, cyan dashed and purple dot-dashed curves, respectively.  }
    \label{fig:sn1987aradialprofiles}
\end{figure}

The radial dependence of temperature $T(r)$ and density $\rho(r)$ on $0<r<R_\nu$ for the three benchmark profiles is shown in the left and right panels of Fig.~\ref{fig:sn1987aradialprofiles}, respectively. As shown in this figure, the Fischer profiles tend to have larger densities than Nakazato 13$M_\odot$, while the temperatures of Fischer 18$M_\odot$ and Nakazato 13$M_\odot$ are expected to be larger than that for Fischer 11.8$M_\odot$. As the production and absorption of $S$ in the supernova core depend both on the baryon density $\rho(r)$ and temperature $T(r)$, all the calculations below will be functions of $r$. To be concrete, we will take the range of $0<r<R_\nu$ for all the three profiles in this paper. This corresponds to the following reasonable simplifications:
\begin{itemize}
    \item Neglecting the production of $S$ beyond the neutrinosphere $R_\nu$.
    \item Assuming the scalar $S$ can stream freely outside the neutrinosphere $R_\nu$ without being absorbed.
\end{itemize}


For simplicity, we assume equivalent numbers of protons and neutrons in the supernova core. This is a reasonable assumption due to the proton and neutron being almost mass degenerate. It should be noted that there could also be some muons in the supernova core, as the ratio $T/m_\mu$ (with $m_\mu$ being the muon mass) is of order 1/3, and hence, not negligible~\cite{Bollig:2020xdr,Croon:2020lrf}. However, the muon number density is roughly one order of magnitude smaller than that for nucleons, and its Yukawa coupling $y_\mu \sim 10^{-3}$ is sufficiently small. Therefore,  we will neglect the muons and focus only on the nucleon bremsstrahlung process for the $S$ production in this paper.

\subsection{Energy loss due to $S$}
\label{sec:sn:calculation}

\begin{figure}[!t]
    \centering
     \includegraphics[width=0.35\textwidth]{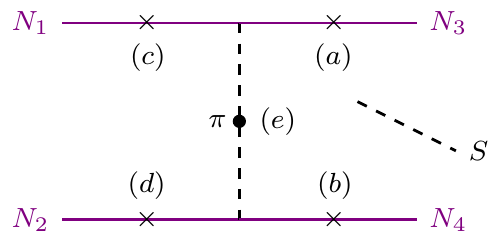}
     \includegraphics[width=0.35\textwidth]{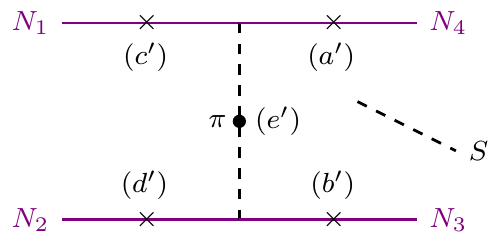}
    \caption{Feynman diagrams for the nucleon-nucleon bremsstrahlung process, with $S$ coupling either to the nucleon legs ($a^{(\prime)}$), ($b^{(\prime)}$), ($c^{(\prime)}$), ($d^{(\prime)}$) or the pion mediator ($e^{(\prime)}$). The left and right panels are respectively the $t$- and $u$-channel diagrams (with $N_3$ and $N_4$ interchanged). }
    \label{fig:diagram:NN}
\end{figure}

Given the mixing of $S$ with the SM Higgs, the scalar $S$ couples to both the nucleons $N = p,\,n$ and the pions $\pi$, with the interaction Lagrangian given by
\begin{eqnarray}
\label{eqn:lagrangian}
{\cal L} \ = \ \sin\theta S
\left[ y_{hNN} \overline{N}N + A_\pi (\pi^0\pi^0 + \pi^+ \pi^-) \right] \,,
\end{eqnarray}
where $y_{hNN} \simeq 10^{-3}$ is the effective coupling of SM Higgs to nucleons~\cite{Shifman:1978zn, Cheng:1988im}, and
\begin{eqnarray}
\label{eqn:Api}
{\cal A}_\pi \ = \ \frac{2}{9v_{\rm EW}}
\left( m_S^2+ \frac{11}{2} m_\pi^2 \right) \,,
\end{eqnarray}
is the effective coupling of $S$ to pions from chiral perturbation theory~\cite{Voloshin:1985tc, Donoghue:1990xh}, with $v_{\rm EW} = (\sqrt2 G_F)^{-1/2} \simeq 246$ GeV the electroweak vacuum expectation value ($G_F$ is the Fermi constant). In the supernova core, the production of scalar $S$ is dominated by the nucleon bremsstrahlung process
\begin{equation}
\label{eqn:brem}
    N + N \to N + N + S \,,
\end{equation}
which is mediated by a single pion at leading order. The corresponding Feynman diagrams are shown in Fig.~\ref{fig:diagram:NN}. Taking into account the couplings of $S$ with both nucleons and the pion mediator in Eq.~(\ref{eqn:lagrangian}), it is found that the contributions from the $S$-nucleon diagrams are partially canceled out. As a result, the matrix element for the process (\ref{eqn:brem}) is dominated by the $SNN$ and $S\pi\pi$ diagrams for the mass ranges of $m_S \gtrsim 10$ MeV and $m_S \lesssim 10$ MeV~\cite{Dev:2020eam, Dev:2021kje}, respectively.

The energy emission rate per unit volume in the supernova core due to the scalar $S$ can be calculated following the standard procedure~\cite{Giannotti:2005tn, Dent:2012mx, Kazanas:2014mca}:
\begin{eqnarray}
\label{eqn:rate}
{\cal Q} (r,\,\phi) \ = \ \int {\rm d} \Pi_5 {\cal S} \sum_{\rm spins} |{\cal M}|^2 (2\pi)^4
\delta^4 (p_1 + p_2 - p_3 - p_4 - k_S) E_S f_1 f_2 P_{\rm decay} P_{\rm abs} \,,
\end{eqnarray}
where $\phi$ is the polar angle for calculating the decay factor $P_{\rm decay}$ and the absorption factor $P_{\rm abs}$ (see Fig.~\ref{fig:geometry}, Eqs.~(\ref{eqn:Rdecay}) and (\ref{eqn:Rabs}) below),  ${\rm d} \Pi_5$ and $\mathcal{M}$ are, respectively, the $2\to 3$ phase space factor and scattering amplitude for the production process (\ref{eqn:brem}), $E_S$ is the energy of $S$, ${\cal S} = 1$ (${1}/{4}$) is the symmetry factor for non-identical (identical) particles in the initial state, and
\begin{eqnarray}
\label{eqn:f12}
f_{1,\,2}({\bf p}; r) \ = \ \frac{\rho(r)}{2m_N}\left(\frac{2\pi}{m_N T(r)}\right)^{3/2}
{\rm{e}}^{-{\bf p}^2/2 m_N T(r)} \,,
\end{eqnarray}
the non-relativistic Maxwell-Boltzmann distributions of the two incoming nucleons in the non-degenerate limit, with density $\rho(r)$ and temperature $T(r)$ both functions of $r$. The integration in Eq.~(\ref{eqn:rate}) can be simplified to~\cite{Dev:2020eam}
\begin{eqnarray}
\label{eqn:rate2}
{\cal Q}(r,\,\phi) & \ = \ & \frac{ \alpha_\pi^2 f_{pp}^4 \sin^2\theta T^{7/2} (r) \rho^2(r)}{8 \pi^{3/2} m_N^{13/2}}\int_q^\infty {\rm d}u \int_0^\infty {\rm d}v \int_{-1}^{1} {\rm d}z
\int_{q}^{\infty} {\rm d}x \ \delta (u-v-x) \nonumber \\
&&\qquad \qquad  \times
\sqrt{uv} e^{-u} x \sqrt{x^2 - q^2} P_{\rm decay} P_{\rm abs} {\cal I}_{\rm tot} \,,
\end{eqnarray}
where $\alpha_\pi \equiv (2m_N/m_\pi)^2/4\pi \simeq 15$, and $f_{pp} \simeq 1$ is the effective pion-nucleon coupling. The dimensionless parameters $u$, $v$, $z$, $x$, $q$ and the dimensionless function ${\cal I}_{\rm tot}$ are collected in Appendix~\ref{sec:appendix}.

\begin{figure}[!t]
    \centering
     \includegraphics[width=0.4\textwidth]{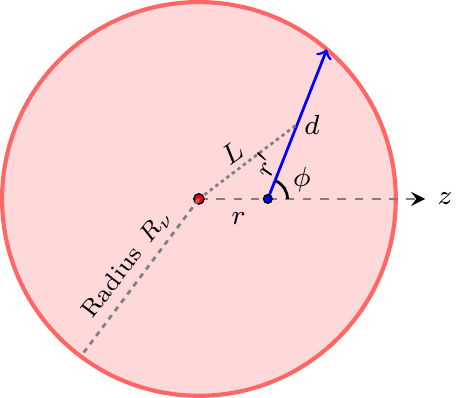}
    \caption{Demonstration of the geometry for the production of $S$ inside the supernova neutrinosphere radius $R_\nu$. The blue line indicates the trajectory of $S$ inside the core (with length $d$), and the geometry is determined by the parameters $r$ and $\phi$. The absorption probability (\ref{eqn:Rabs}) depends on the distance $L$ which is a function of $r'$ in the range of $0<r'<d$.  }
    \label{fig:geometry}
\end{figure}

Na\"{i}vely speaking, only $S$ decaying outside the neutrinosphere $R_\nu$ can contribute to energy loss, which is measured by the factor $P_{\rm decay}$ in Eq.~(\ref{eqn:rate}). However, this depends on the geometric factor mentioned in the introduction, which is related to the production site of $S$ in the core and the flight direction of $S$. As illustrated in Fig.~\ref{fig:geometry}, this can be determined by two parameters:
\begin{itemize}
    \item Assuming the supernova core is spherically symmetric, the production site of $S$ is determined solely by the distance $r$ to the center.
    \item The direction of the momentum of $S$ can be characterized by the ``polar angle'' $\phi$ between the 3-momentum of $S$ and the $z$-direction, where the $z$-direction is defined by connecting the production site and the stellar center. The angle $\phi$ will be relevant when one integrates over the phase space of $S$ (cf. Eq.~(\ref{eqn:LS})).
\end{itemize}
As shown in Fig.~\ref{fig:geometry}, the length $d$ of the trajectory $S$ travels inside the star is a function of $r$ and $\phi$ via
\begin{align}
    d(r,\phi)=\sqrt{R_\nu^2-r^2 \sin^2\phi} - r\cos\phi \,.
\end{align}
Then the decay probability is
\begin{eqnarray}
\label{eqn:Rdecay}
P_{\rm decay} (r, \phi) \ = \ {\rm exp} \Big\{ -d(r,\phi) \Gamma_S \Big\} \,,
\end{eqnarray}
where $\Gamma_{S} = (m_S/E_S)\Gamma_{0,\,S}$, with $E_S/m_S$ the Lorentz boost factor, and $\Gamma_{0,\,S}$ the proper total decay width of $S \to e^+ e^-,\, \mu^+ \mu^-,\, \pi^+ \pi^-,\, \pi^0 \pi^0,\, \gamma\gamma$, which can be found in Ref.~\cite{Dev:2020eam}. We have assumed that the decay products within the neutrinosphere $R_\nu$ do not contribute to the cooling and immediately thermalize~\cite{Dent:2012mx}. As the mixing parameter $\sin\theta$ becomes large, the decay of $S$ happens earlier after its production and the scalars are not effective at cooling, due to exponential modulation of the emission rate. The factor $P_{\rm abs}$ in Eq.~(\ref{eqn:rate}) accounts for the absorption of $S$ inside the star due to the inverse nucleon bremsstrahlung process $N + N + S \ \to \ N + N$, which can be written as
\begin{eqnarray}
\label{eqn:Rabs}
P_{\rm abs} (r,\phi) \ = \ {\rm exp} \left\{
- \int_0^d  \frac{{\rm d}r^\prime}{\lambda[L(r,\phi;r^\prime)]} \right\} \,,
\end{eqnarray}
with $\lambda$ the MFP of $S$, which is function of the length $L$ in Fig.~\ref{fig:geometry} when $S$ travels through the star. $L$ is not only a function of $r$ and $\phi$, but also depends on the position of $S$ along the distance $d$ in Fig.~\ref{fig:geometry}:
\begin{eqnarray}
L (r,\phi;r^\prime) = \sqrt{r^2 + r^{\prime 2} + 2 r r^\prime \cos\phi} \,,
\end{eqnarray}
as $L$ varies when the auxiliary parameter $r^\prime$ changes from $0$ to $d$. Therefore, we integrate over $r^\prime$ in between $0$ and $d$ in Eq.~(\ref{eqn:Rabs}). In the case of constant MFP $\lambda$, the absorption factor simply reduces to ${\rm exp} \left\{ - d/\lambda \right\}$.
For the supernova profiles adopted in this paper, the MFP
$\lambda$ depends on the density $\rho(r)$ and temperature $T(r)$ in Fig.~\ref{fig:sn1987aradialprofiles}. For a sufficiently large coupling, the particle $S$ can be trapped within the supernova, thus does not contribute to the energy loss. This provides a limit on the coupling, above which the trapping does not allow us to put a supernova constraint on the $S$ particle.

The inverse MFP of $S$ for the inverse bremsstrahlung process is given by~\cite{Burrows:1990pk, Giannotti:2005tn}
\begin{eqnarray}
\label{eqn:mfp}
\lambda^{-1}(r;x) \ = \
\frac{1}{2 E_S}
\int {\rm d} \Pi_4 {\cal S} \sum_{\rm spins} |{\cal M}|^2 (2\pi)^4
\delta^4 (p_1 + p_2 - p_3 - p_4 + k_S) f_1 f_2 \,,
\end{eqnarray}
where ${\rm d}\Pi_4$ is the four-body phase space for the initial and final state nucleons. It is clear that the MFP $\lambda(r;x)$ depends on the scalar energy $E_S$, or equivalently on the dimensionless parameter $x \equiv E_S/T$. The simplified expression of $\lambda^{-1}$ in terms of the dimensionless parameters and functions in Appendix~\ref{sec:appendix} is
\begin{eqnarray}
\label{eqn:mfp2}
\lambda^{-1} (r;x) & \ = \ &
\frac{\pi^{1/2} \alpha_\pi^2 f_{pp}^4 \sin^2\theta \rho^2(r)}{4m_N^{13/2} T^{1/2}(r)} \frac{1}{x}
\int_0^\infty {\rm d}u \int_q^\infty {\rm d}v \int_{-1}^{1} {\rm d}z \sqrt{uv} e^{-u}
\delta (u-v+x) {\cal I}_{\rm tot} \,. \nonumber \\ &&
\end{eqnarray}
As expected, due to the factor $1/x$ in the equation above, the more energetic scalars tend to have a larger MFP. Taking this into account, we will use the energy-dependent $\lambda(r;x)$ instead of the energy-independent MFP $\langle \lambda \rangle (r)$ for the calculation of ${\cal Q}$, which is an improvement over the simplified treatment in Ref.~\cite{Dev:2020eam}.

\begin{figure}[!t]
    \centering
     \includegraphics[width=0.48\textwidth]{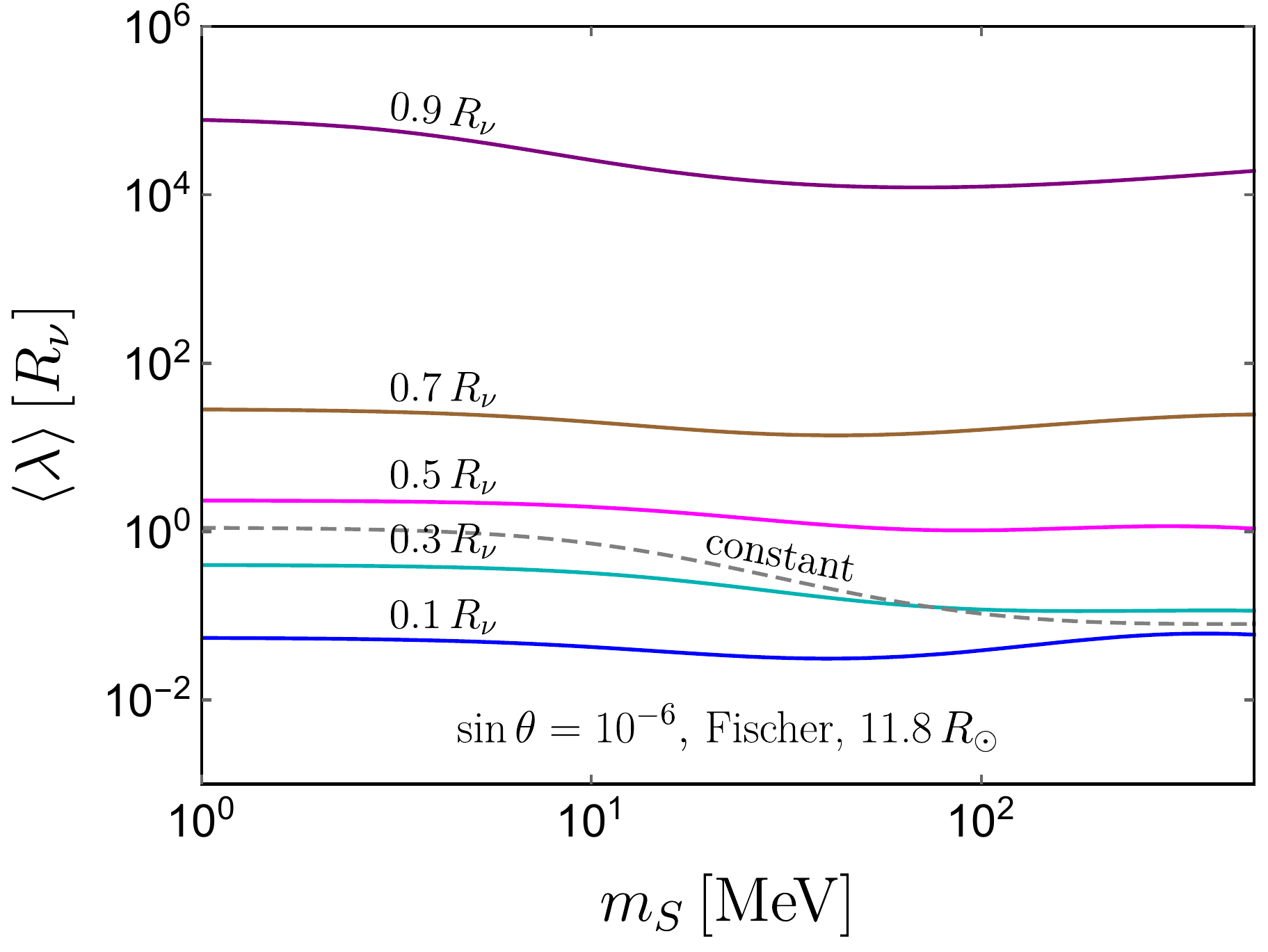}     \includegraphics[width=0.48\textwidth]{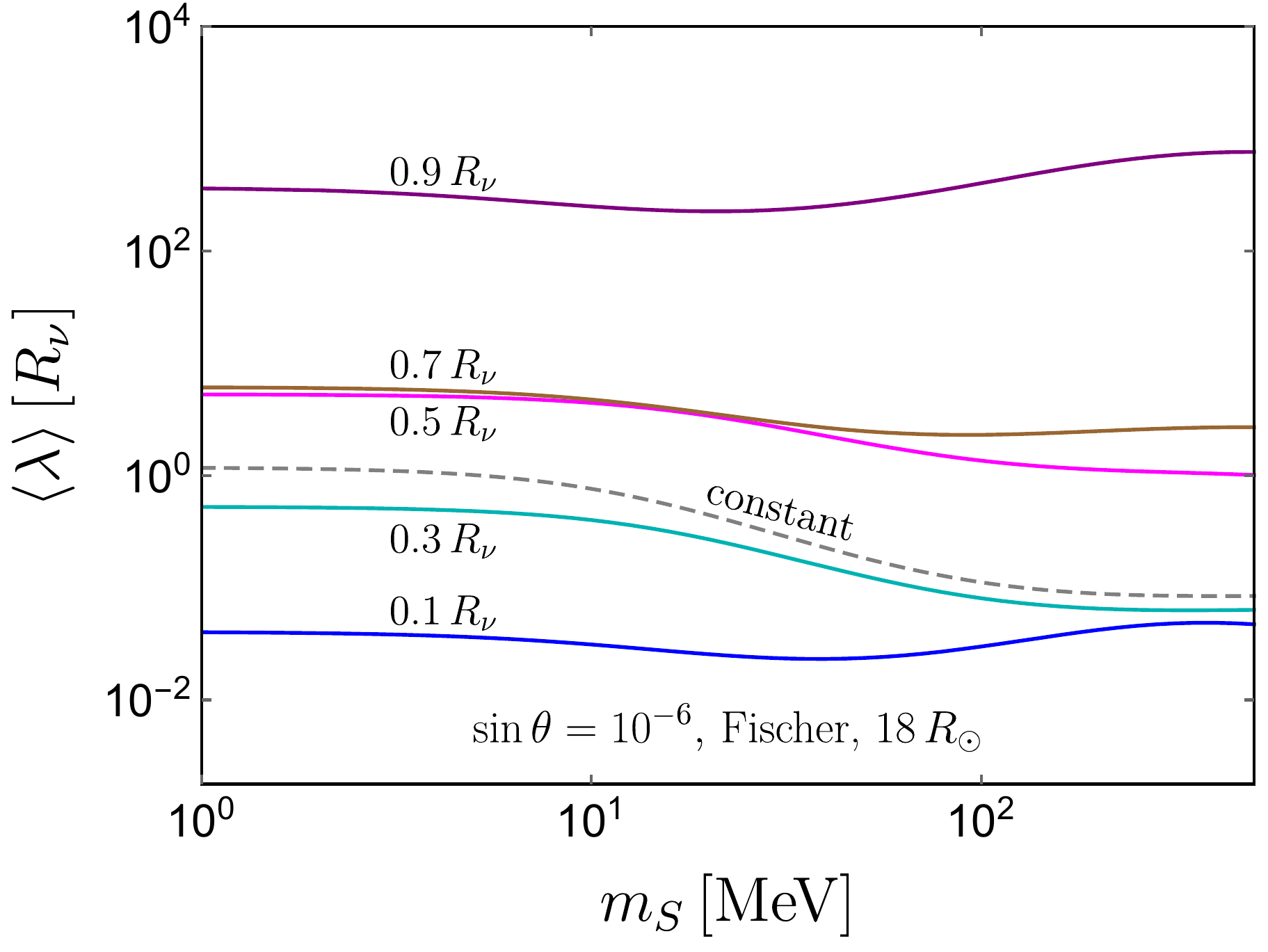}\\
    \includegraphics[width=0.48\textwidth]{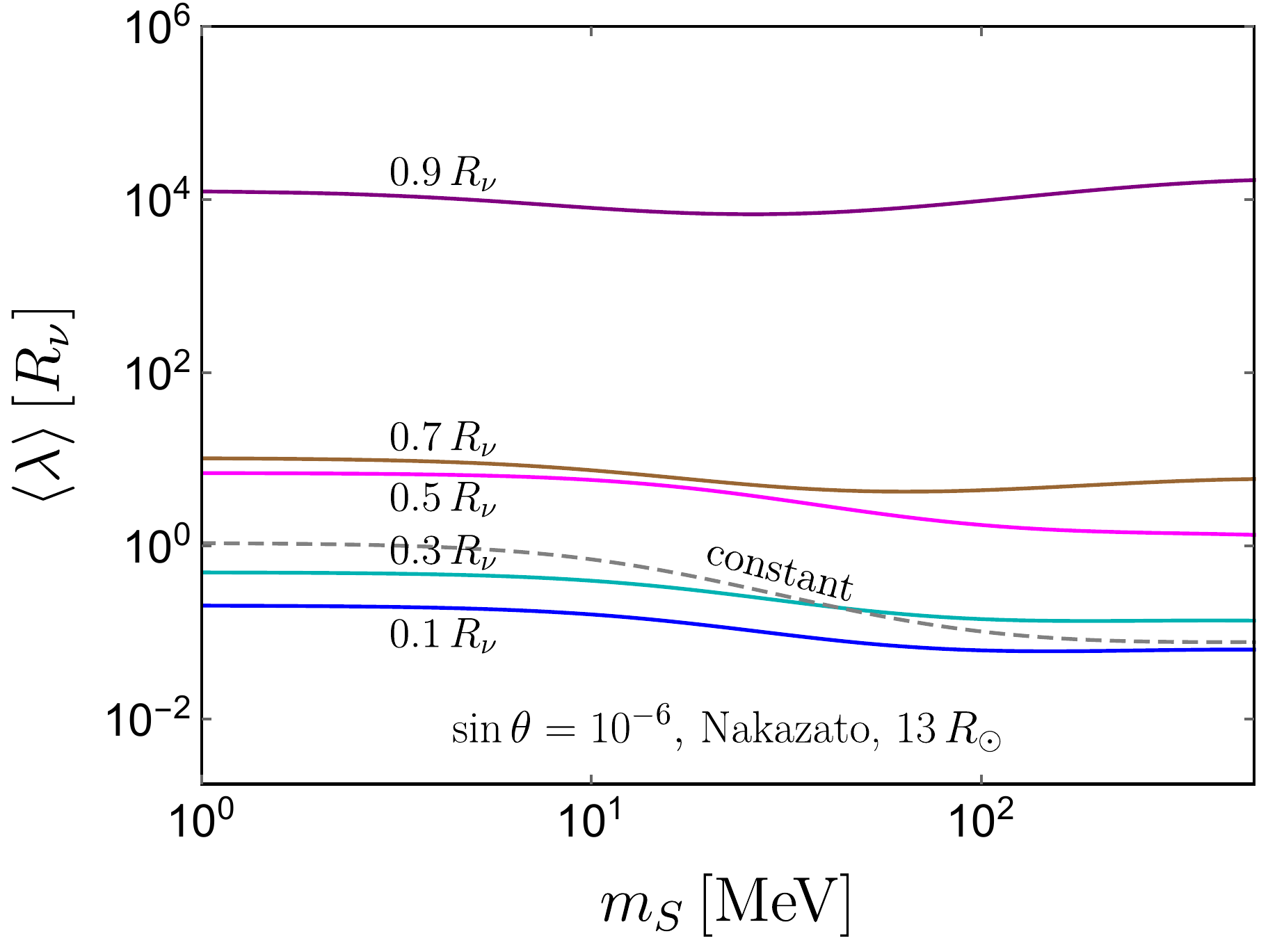}
    \caption{The MFP $\langle \lambda \rangle$ of $S$ in the supernova core as function of $m_S$ at the radial positions $r=[0.1,\, 0.3,\, 0.5,\, 0.7,\, 0.9]R_\nu$, for the SN1987A profiles Fischer 11.8$M_\odot$ (upper left), Fischer 18$M_\odot$ (upper right) and Nakazato 13$M_\odot$ (bottom). The dashed lines are the corresponding MFPs for constant temperature $T= 30$ MeV and baryon number density $n_B = 1.2 \times 10^{38} \, {\rm cm}^{-3}$~\cite{Dev:2020eam}.  The mixing angle is fixed to $\sin\theta=10^{-6}$. Note that the radius $R_\nu$ is different in the three supernova profiles (cf.~Table~\ref{tab:summary}).}
    \label{fig:snmfp}
\end{figure}

To demonstrate the dependence of MFP on the radius $r$ and supernova profiles, we resort to the effective energy-independent MFP $\langle \lambda \rangle (r)$, with its inverse defined by~\cite{Ishizuka:1989ts}
\begin{eqnarray}
\label{eqn:lambda:average}
\langle \lambda^{-1} \rangle (r) \ \equiv \
\frac{\bigintss {\rm d} E_S \frac{E_S^3}{e^{E_S/T}-1} \lambda^{-1} (E_S; r)}{\bigintss {\rm d} E_S \frac{E_S^3}{e^{E_S/T}-1}}
\ = \ \frac{\bigintss {\rm d} x \frac{x^3}{e^{x}-1} \lambda^{-1} (x;r)}{\bigintss {\rm d} x \frac{x^3}{e^{x}-1}} \,.
\end{eqnarray}
The MFPs $\langle \lambda \rangle$ of $S$ in the supernova core at various radial positions $r$ = [0.1, 0.3, 0.5, 0.7, 0.9] $R_\nu$ are shown in Fig.~\ref{fig:snmfp} as functions of the scalar mass $m_S$, where the upper left, upper right and lower panels are for the Fischer $11.8M_\odot$, Fischer $18M_\odot$ and Nakazato 13$M_\odot$ profiles, respectively. The dashed gray lines are the corresponding MFPs for the constant temperature $T = 30$ MeV and baryon number density $n_B = 1.2 \times 10^{38} \, {\rm cm}^{-3}$ in the supernova core from Ref.~\cite{Dev:2020eam}. To be explicit, we have set the mixing angle to $\sin\theta = 10^{-6}$. For other values of mixing angle, the MFP simply scales as $\langle \lambda \rangle \propto \sin^{-2}\theta$. As shown in Fig.~\ref{fig:snmfp}, in the inner core $r \sim 0.3 R_\nu$, the MFP $\langle \lambda \rangle$ is close to that in the case of constant temperature and baryon density, both at the order of $R_\nu \sim {\cal O} (20\, {\rm km})$. In the outer layers, say $r \gtrsim 0.5 R_\nu \sim 10$ km, both temperature and density drop significantly (cf.~Fig.~\ref{fig:sn1987aradialprofiles}), and the MFP grows quickly when $r$ approaches $R_\nu$. As indicated by the purple lines in Fig.~\ref{fig:snmfp},  the MFPs at $r = 0.9 R_\nu$ in the three supernova profiles are orders of magnitude larger than that for $r \sim 0.3 R_\nu$. In other words, the scalars produced in the outer regions of the supernova core have a much higher chance of escaping than those produced in the inner regions. This large hierarchy of MFP as a function of radial position has been taken into account by the integration in Eq.~(\ref{eqn:Rabs}).

\subsection{Results}
\label{sec:sn:results}

Integrating over the whole volume of SN1987A inside $R_\nu$, we arrive at the luminosity due to the emission of the scalar $S$ via the expression
\begin{eqnarray}
\label{eqn:LS}
{\cal L}_S = \int {\cal Q}(r,\phi) {\rm d}V
= 2\pi \int_0^{R_\nu} {\rm d}r \, r^2 \int_0^{\pi} {\rm d}\phi \sin\phi \, {\cal Q}(r,\phi) \,.
\end{eqnarray}
The dependence on the ``polar angle'' $\phi$ is from the decay probability $P_\textrm{abs}$ in Eq.~(\ref{eqn:Rdecay}) and the absorption probability $P_\textrm{decay}$ in Eq.~(\ref{eqn:Rabs}).
With respect to the constant density and temperature case in Ref.~\cite{Dev:2020eam}, the improvements in this paper are:
\begin{itemize}
    \item The supernova profiles in Fig.~\ref{fig:sn1987aradialprofiles} include the radial dependence of density $\rho(r)$ and temperature $T(r)$. These were assumed to be constants in Ref.~\cite{Dev:2020eam}. This will affect all the consequent results, for instance the MFP $ \lambda(r;x)$, as discussed above.

    \item The geometric effect in Fig.~\ref{fig:geometry} for the decay and absorption of $S$ in the supernova core has been taken into account. In Ref.~\cite{Dev:2020eam}, the decay length and MFP $\langle \lambda \rangle$ were compared na\"{i}vely with the radius $R_c$ of the supernova core.

    \item The MFP $\lambda(r;x)$ is adopted for the calculation of ${\cal Q}$ to take into account the energy dependence of MFP, whereas the effective energy-independent MFP $\langle \lambda \rangle$ is used in Ref.~\cite{Dev:2020eam}.
\end{itemize}
With these factors taken into consideration, the integration in the master formula (\ref{eqn:LS}) becomes much more complicated and computationally expensive, although the calculations are in some sense straightforward.

\begin{figure}[!t]
    \centering
     \includegraphics[height=0.24\textheight]{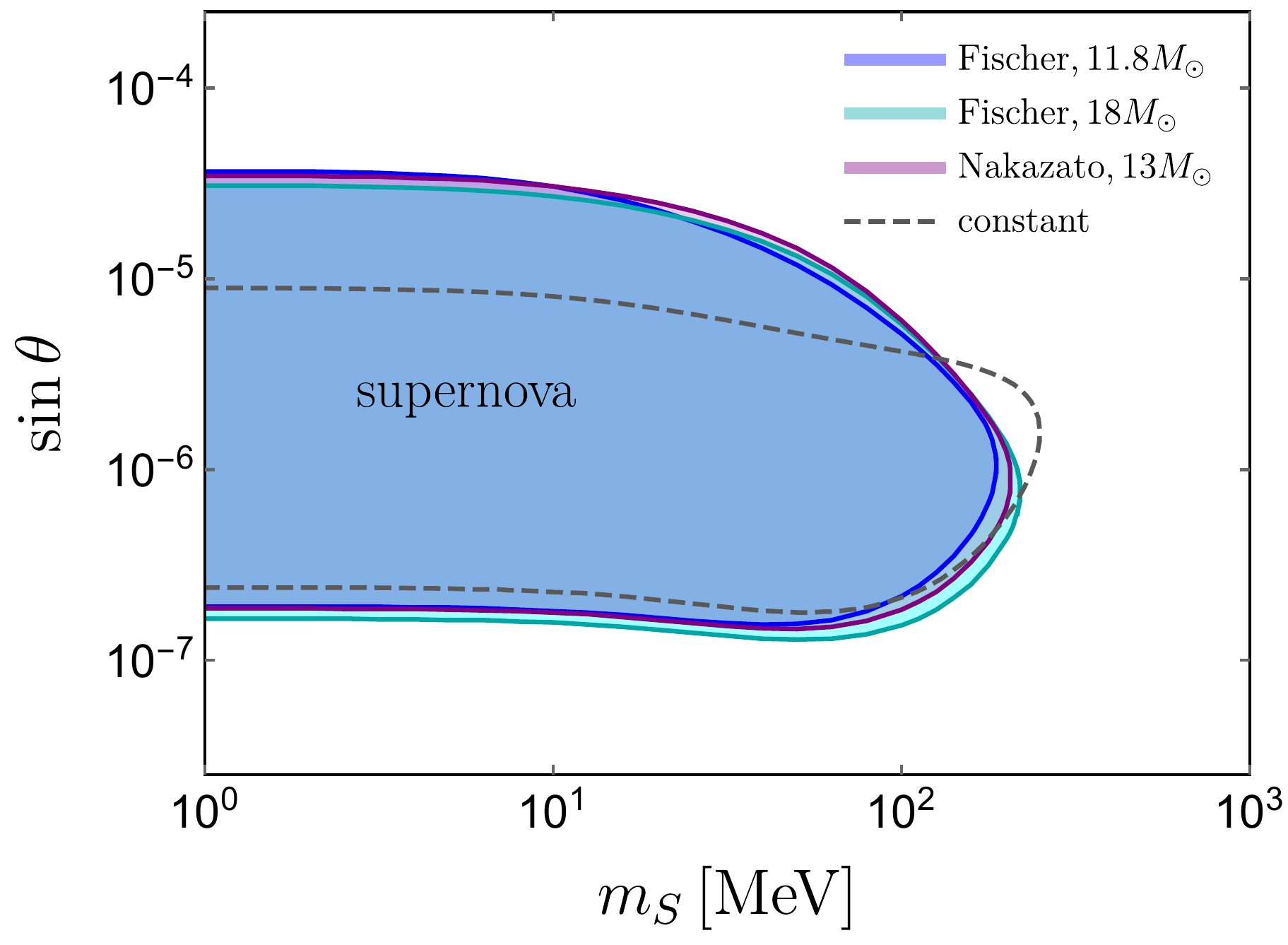}
     \includegraphics[height=0.24\textheight]{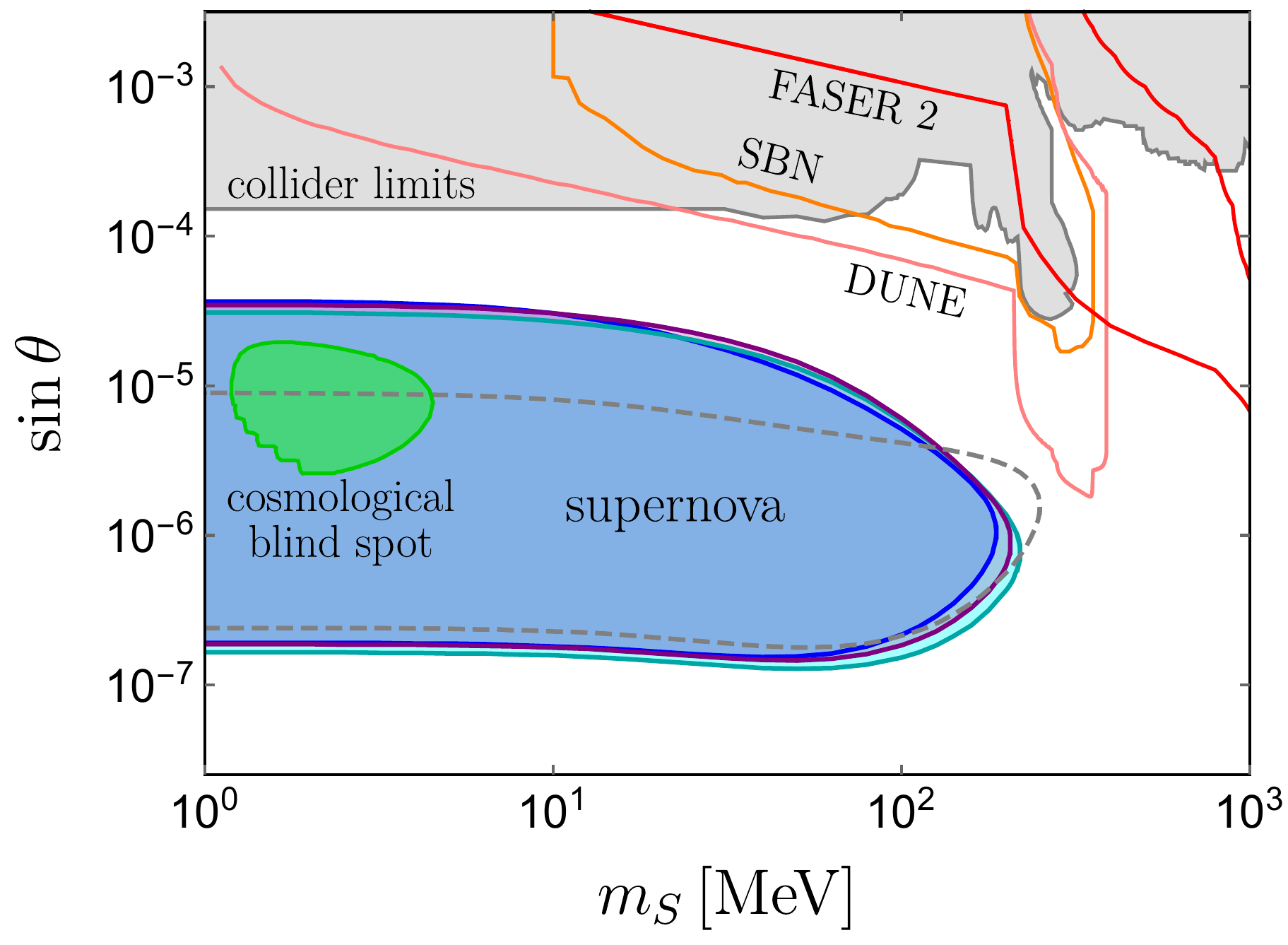}
    \caption{{\it Left panel}: SN1987A limits on the scalar mass $m_S$ and mixing angle $\sin\theta$, assuming Fischer 11.8$M_\odot$ (blue), Fischer 18$M_\odot$ (cyan) and Nakazato 13$M_\odot$ (purple) supernova profiles. Also shown is the limit in dashed gray line with constant density $n_B = 1.2 \times 10^{38}\ {\rm cm}^{-3}$ and temperature $T = 30$ MeV within core radius $R_c=10$ km. All these bounds assume that the luminosity ${\cal L}_S < 3\times 10^{52}$ erg/sec. {\it Right panel}: Complementarity of the supernova limits with those from collider searches (shaded gray)~\cite{Dev:2017dui, Egana-Ugrinovic:2019wzj, Dev:2019hho} and the cosmological blind spot with a high reheating temperature of $\gtrsim100$ GeV (shaded green)~\cite{Ibe:2021fed}. The orange, pink and red lines indicate the future prospects of $S$ at SBN~\cite{Batell:2019nwo}, DUNE~\cite{Berryman:2019dme} and FASER 2~\cite{Anchordoqui:2021ghd}. More cosmological limits can be found in e.g. Refs.~\cite{Berger:2016vxi, Fradette:2018hhl, Ibe:2021fed}. }
    \label{fig:snlimits}
\end{figure}

To set limits on the scalar mass $m_S$ and the mixing angle $\sin\theta$, we conservatively require that the luminosity ${\cal L}_S$ is smaller than 10\% of the measured neutrino luminosity ${\cal L}_\nu \simeq 3\times 10^{53}$ erg/sec, so
\begin{eqnarray}
\label{eqn:sn:limit}
{\cal L}_S < 3 \times 10^{52} \ {\rm erg/sec} \,.
\end{eqnarray}
The resultant supernova limits on $m_S$ and $\sin\theta$ for the profiles Fischer $11.8M_\odot$, Fischer $18M_\odot$ and Nakazato 13$M_\odot$ are shown in the left panel of Fig.~\ref{fig:snlimits}, as the blue, cyan and purple shaded regions, respectively. For comparison, the limit for the constant density and temperature case is shown as the gray contour, for which the geometry in not considered, and we take $n_B = 1.2 \times 10^{38}\ {\rm cm}^{-3}$ and $T = 30$ MeV, $R_c = 10$ km, and the luminosity limit in Eq.~(\ref{eqn:sn:limit}).
The following features are observed for the updated supernova limits:
\begin{itemize}
    \item The whole volume $V\propto R_\nu^3$ within the neutrinosphere is significantly larger than $V_c \propto R_c^3$ by a factor of $\sim (25\, {\rm km} / 10\, {\rm km})^3 \simeq 15$, though the density and temperature are lower in the outer layers (cf. Fig.~\ref{fig:sn1987aradialprofiles}). With respect to the constant profile case, more $S$ can be produced in all three supernova profiles adopted in this work. Mixing angles down to $1.5\times10^{-7}$, $1.3\times10^{-7}$ and $1.5\times10^{-7}$ are excluded for the profiles Fischer 11.8$M_\odot$, Fischer 18$M_\odot$ and Nakazato 13$M_\odot$, respectively, which are better than the constant profile case of $1.8\times10^{-7}$ by a factor of 1.2, 1.4 and 1.2, respectively.
    \item As presented in Fig.~\ref{fig:snmfp}, the MFP is (much) longer for all three supernova profiles, than that in the constant profile case when $r\gtrsim0.5 R_\nu$. Furthermore, without averaging over the energy dependence of MFP, more energetic scalars tend to have larger MFP. It turns out  that the overall absorption effect is weaker when the profiles are adopted, and more light scalars produced can escape from the core and contribute to energy loss. As a result, a larger mixing angle $\sin\theta$ can now be excluded for all  three profiles. These limits turn out to be $3.8\times10^{-5}$, $3.1\times10^{-5}$ and $3.6\times10^{-5}$ for the profiles Fischer 11.8$M_\odot$, Fischer 18$M_\odot$ and Nakazato 13$M_\odot$ at $m_S = 1$ MeV, improving the constant profile limit of $9.0\times10^{-6}$ by a factor of 4.2, 3.5 and 4.0, respectively.
    \item With the profiles Fischer 11.8$M_\odot$, Fischer 18$M_\odot$ and Nakazato 13$M_\odot$, the scalar mass ranges up to, respectively, 187 MeV, 219 MeV and 205 MeV are excluded, which are close to the constant profile case of 249 MeV. On the other hand, the supernova limits in Fig.~\ref{fig:snlimits} extend to the mass range of $m_S <$ MeV, even down to the massless limit $m_S \to 0$. However, such light scalars are stringently constrained by the cosmological limits~\cite{Berger:2016vxi, Fradette:2018hhl, Ibe:2021fed}.
\end{itemize}

The supernova limits on $S$ are largely complementary to those from collider searches, cosmological observations and other astrophysical limits, as demonstrated in the right panel of Fig.~\ref{fig:snlimits}.
From mixing with the SM Higgs field, the scalar $S$ can have flavor-changing neutral-current (FCNC) couplings to the SM quarks at the 1-loop level. Therefore, for $m_S \lesssim$ GeV, $S$ can be produced from FCNC decays of SM mesons, e.g. $K,\,B \to \pi + S$ and $B \to K + S$. Then the scalar mass $m_S$ and mixing angle $\sin\theta$ can be constrained by the high-precision laboratory meson data, e.g. those from  NA48/2~\cite{Batley:2009aa,Batley:2011zz}, E949~\cite{Artamonov:2009sz}, KOTO~\cite{Ahn:2018mvc}, NA62~\cite{Ceccucci:2014oza,Ruggiero:2019, NA62}, KTeV~\cite{AlaviHarati:2003mr,AlaviHarati:2000hs,Alexopoulos:2004sx, Abouzaid:2008xm}, BaBar~\cite{Aubert:2003cm,Lees:2013kla}, Belle~\cite{Wei:2009zv}, LHCb~\cite{Aaij:2012vr, LHCb:2015nkv, LHCb:2016awg} and CHARM~\cite{Bergsma:1985qz}. The scalar $S$ can also be produced at colliders via the proton bremsstrahlung process, and the most stringent limit in such channels is from LSND~\cite{Foroughi-Abari:2020gju}. Combining all these laboratory limits, the mixing angle $\sin\theta \gtrsim 10^{-4}$ is excluded for $m_S \lesssim 300$ MeV~\cite{Dev:2017dui, Egana-Ugrinovic:2019wzj, Dev:2019hho}. These current collider limits are shown as  the gray shaded region in the right panel of Fig.~\ref{fig:snlimits}. At future high-intensity experiments, more parameter space of $m_S$ and $\sin\theta$ can be probed, e.g. at SBN~\cite{Batell:2019nwo}, DUNE~\cite{Berryman:2019dme, Dev:2021qjj} and FASER 2~\cite{Anchordoqui:2021ghd}, which are presented in the right panel of Fig.~\ref{fig:snlimits} as the orange, pink and red curves, respectively. More details of the collider limits can be found e.g. in Ref.~\cite{Dev:2017dui, Egana-Ugrinovic:2019wzj, Dev:2019hho}.

For the scalar $S$, neutron star (NS) mergers present a similar environment to supernovae, with potentially larger density and temperature. Therefore $S$ can also be produced in mergers via the nucleon bremsstrahlung process. Depending on the scalar mass $m_S$ and the mixing angle $\sin\theta$ as well as the density and temperature, the scalar $S$ may be trapped in or free stream from the merger~\cite{Dev:2021kje}. For the free streaming regions, the scalar $S$ might take away sizable energy from the stars, thus contributing to energy loss. In the trapped regions, $S$ will provide a new mechanism for thermal conduction. In some regions of the parameter space, the thermal conductivity of $S$ may even surpass that coming from trapped neutrinos, thus significantly affecting the merger evolution. However, there is not enough observational data at this stage to put a meaningful merger constraint on the $S$ parameter space shown in Fig.~\ref{fig:snlimits}, although this has promising prospects in the future~\cite{Dev:2021kje}.

In the early Universe, a light scalar $S$ with very small couplings to the SM particles contributes to the light degrees of freedom $N_{\rm eff}$ and spoils the successful big bang nucleosynthesis (BBN), depending on the reheating temperature $T_R$. For the mixing angle in the range of interest for the supernova limits, with $T_R = 5$ MeV, the limit is $m_S > 1$ MeV. With $T_R \gtrsim 100$ GeV, the cosmological limit is very sensitive to the mixing angle $\sin\theta$, excluding scalar mass even above GeV~\cite{Berger:2016vxi, Fradette:2018hhl, Ibe:2021fed}. However, there is a ``blind spot'' left unconstrained at $m_S \sim 2$ MeV and $\sin\theta \sim 10^{-5}$~\cite{Ibe:2021fed}, which is labeled as the shaded green region in the right panel of Fig.~\ref{fig:snlimits}. It is remarkable that the cosmological ``blind spot'' is well excluded by the supernova limits with the three profiles adopted in this paper, which is otherwise only partially excluded if the constant profile case is considered for supernova limits. However, one should note that the cosmological limits on $S$ depend not only on the self-interactions of $S$, e.g. the quartic coupling $\lambda_S S^4/4!$, but also on the statistical treatment of the cosmological microwave background (CMB) limit~\cite{Ibe:2021fed}. Furthermore, with a non-standard history, the cosmological limits on $S$ might also be dramatically altered~\cite{Bernal:2018kcw}. Therefore, the astrophysical limits derived here should be treated as independent probes of the scalar parameter space, irrespective of the cosmological bounds.

\section{$S$ in the Sun}
\label{sec:Sun}

\subsection{Standard solar model}
\label{sec:sun:ssm}

\begin{figure}[!t]
    \centering
    \includegraphics[height=0.23\textheight]{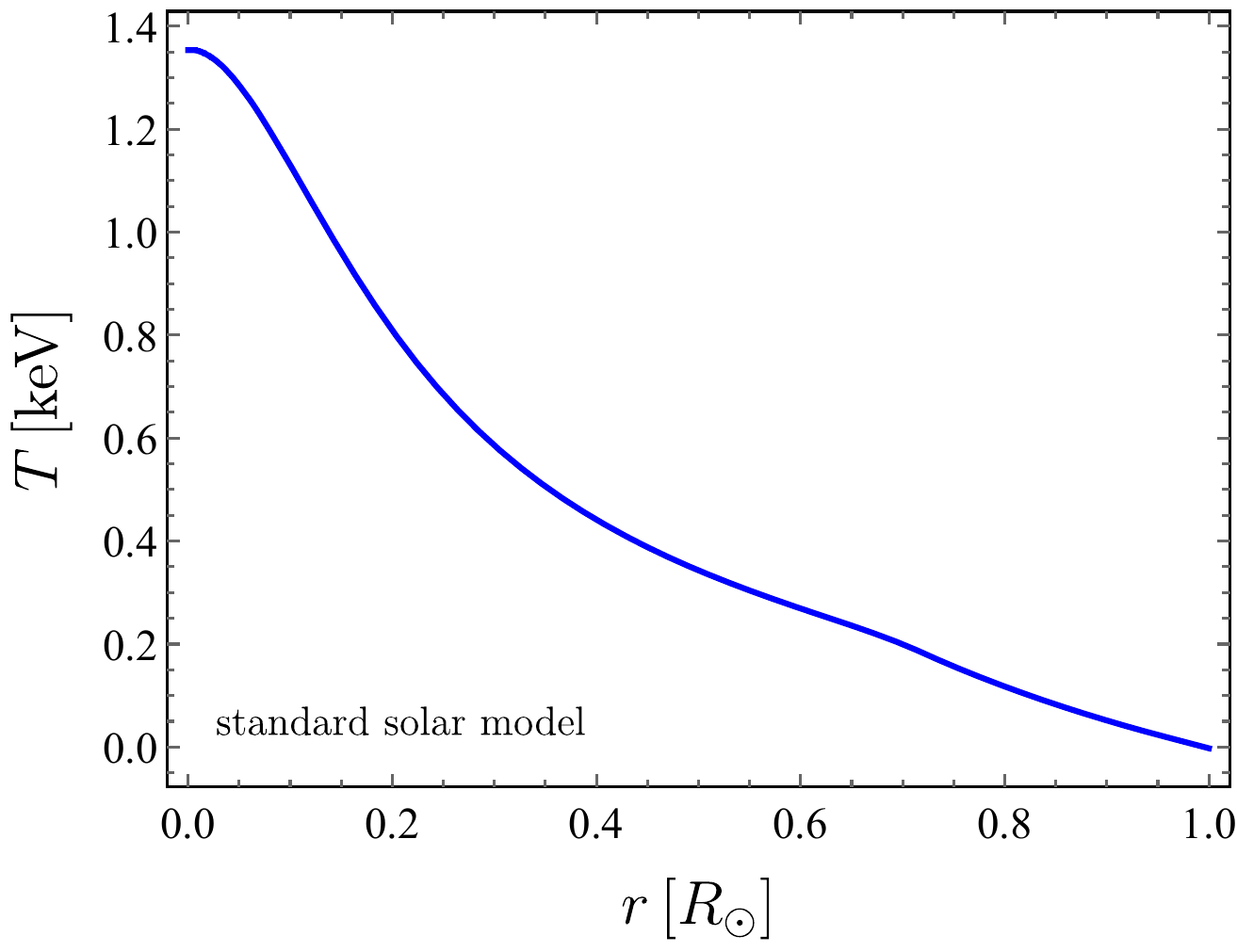}
    \includegraphics[height=0.23\textheight]{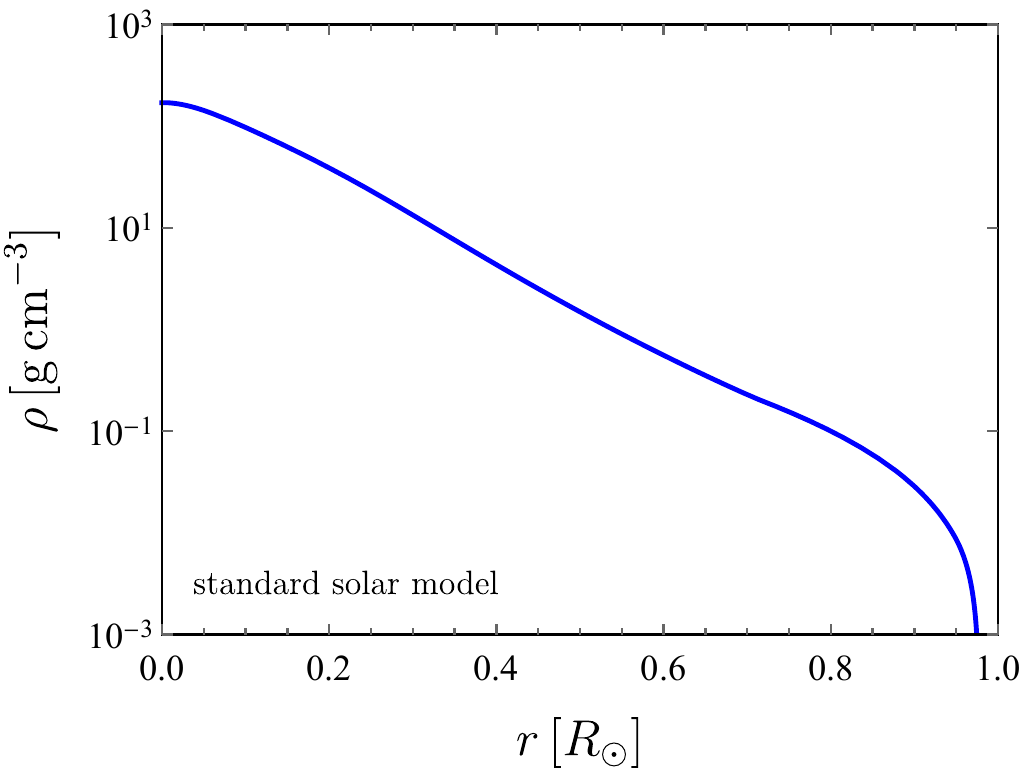}\\
    \includegraphics[height=0.23\textheight]{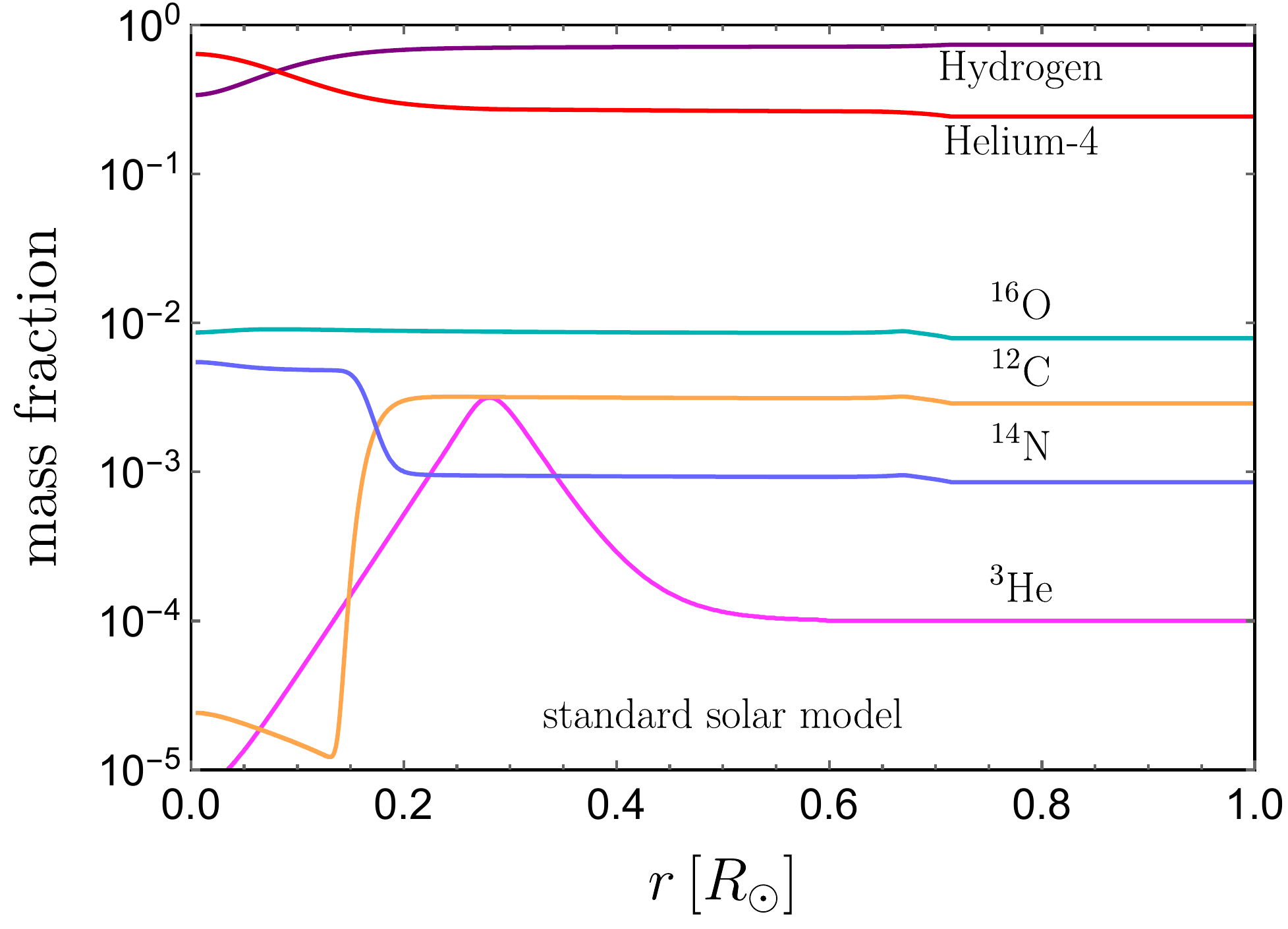}
    \caption{The profiles for temperature $T$ (upper left), density $\rho$ (upper right) and mass fractions of the elements Hydrogen, Helium-4, $^3$He, $^{12}$C, $^{14}$N and $^{16}$O (lower), as functions of the radial position $r$ for the standard solar model.  }
    \label{fig:solarprofies}
\end{figure}

For studying limits obtainable with the Sun, we adopt the standard solar profile~\cite{Bahcall:2004fg}. In this model, the Sun is modeled as a spherically symmetric quasi-static star. The stellar structure is described completely by a set of differential equations and boundary conditions for the luminosity, radius, age and composition of the Sun, which are well understood~\cite{Bahcall:2004fg, Bahcall:2004pz, Bahcall:2005va}. The differential equations involve pressure, opacity and the energy generation rate  written in terms of the density, temperature and composition. It notably provides precise estimates for the Helium abundance and mixing length parameter, which are computed in a definite manner by fitting the stellar model to the observed solar luminosity and radius at the Sun's present age. In the solar center, the temperature $T_\odot(r)$ can exceed 1.2 keV, and the density $\rho_\odot(r)$ can reach up to 150 g/cm$^3$. The temperature $T_\odot(r)$ drops gradually while $\rho_\odot(r)$ gets suppressed very quickly as $r$ approaches the solar radius $R_\odot \simeq 7.0\times10^5$ km, as shown in the upper panels of Fig.~\ref{fig:solarprofies}. The Sun is composed of mostly Hydrogen and Helium-4 ions, and their mass fractions are 34\% and 64\% at the solar core, changing smoothly to 74\% and 24\% at the solar surface, respectively, as shown in the lower panel of Fig.~\ref{fig:solarprofies}. The remaining $2\%$ mass fraction is Helium-3 and other heavy elements such as $^{12}$C, $^{14}$N and $^{16}$O.

\subsection{Energy loss due to $S$}
\label{sec:sun:production}

\begin{figure}[!t]
    \centering
     \includegraphics[height=0.12\textheight]{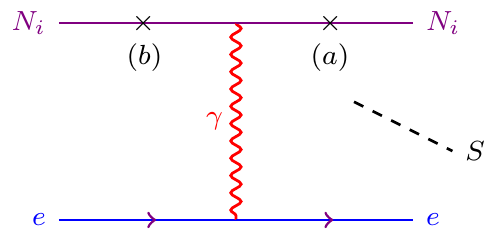}
    \caption{Feynman diagram for the dominant production mechanism of $S$ in the Sun via the  $e-N_i$ bremsstrahlung process, where the scalar $S$ can couple to either of the $N_i$ fermion legs $(a)$ or $(b)$. }
    \label{fig:diagram:sun}
\end{figure}

\begin{figure}[!t]
    \centering
     \includegraphics[width=0.6\textwidth]{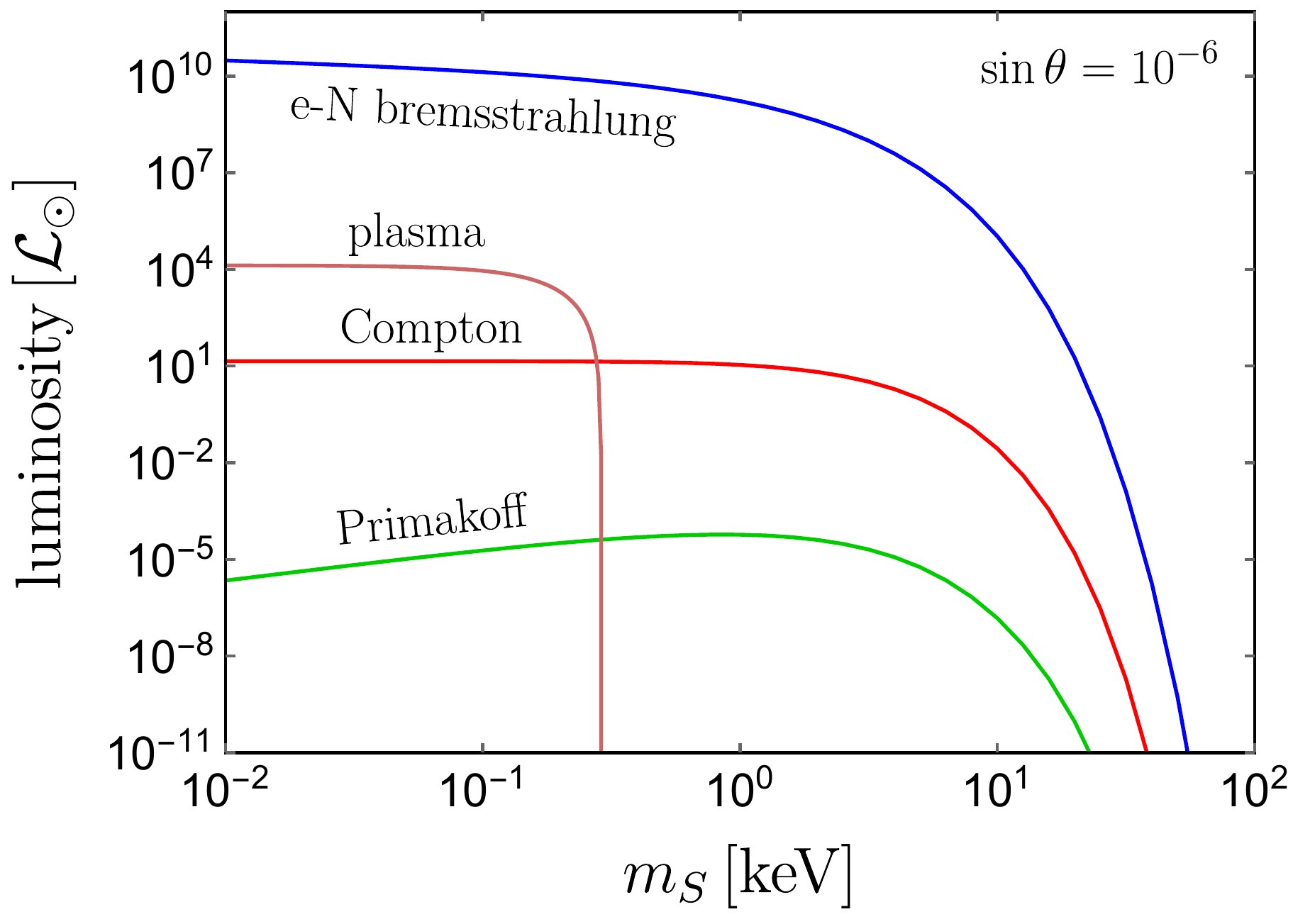}
    \caption{Luminosities due to the different production channels of $S$ in the Sun: $e-N$ bremsstrahlung, Compton and Primakoff processes and the plasma contribution, as functions of the scalar mass $m_S$. The solar profile is taken into account, but the decay and absorption of $S$ inside the Sun are not considered in this plot. The scalar mixing is fixed to be $\sin\theta=10^{-6}$. }
    \label{fig:channels}
\end{figure}

The solar temperature is at the keV scale, and the dominant production channel of $S$ in the Sun is via the $e-N_i$ bremsstrahlung process, which is mediated by a photon, with the scalar $S$ coupling to the nucleus $N_i$. The corresponding Feynman diagram is shown in Fig.~\ref{fig:diagram:sun}. The scalar can also couple to the electron and the photon mediators, however, these couplings are highly suppressed by the small electron Yukawa coupling $y_e$ and the loop-level coupling of $S$ with photons respectively. The scalar $S$ can also be produced from the $e-e$ and $N_i - N_i$ bremsstrahlung processes, the Compton-like process $e+\gamma \to e + S$, the Primakoff process $\gamma + X \to X + S$ (with $X$ being electron or the nuclei). However, these channels are comparatively much smaller~\cite{Dev:2020jkh}.
The plasma effect can also contribute to the production of $S$, due to the mixing of $S$ with the longitudinal massive photon in the media. However, it is found that the plasma effect is subdominant relative to the $e-N_i$ bremsstrahlung process~\cite{Dev:2020jkh}. The comparison of luminosities due to these channels is presented in Fig.~\ref{fig:channels}, where we have set the mixing angle to be $\sin\theta = 10^{-6}$. The decay and absorption of $S$ in the Sun are not taken into consideration; however, different from Ref.~\cite{Dev:2020jkh}, we have included the solar profile in the calculations of luminosities. It is clear in  Fig.~\ref{fig:channels} that the luminosity in the $e-N$ bremsstrahlung process is orders of magnitude larger than all other channels. In the following sections we will consider only the dominant production channel, and the emission rates in the subdominant channels are collected in Appendix~\ref{sec:appendix2}.



The energy emission rate per unit volume in the Sun due to the $e-N_i$ bremsstrahlung process is given by
\begin{eqnarray}
\label{eqn:rate:master}
{\cal Q}_{}^{} (r,\phi) & = & \sum_i
\int {\rm d} \Pi_5 \sum_{\rm spins} |{\cal M}_i|^2 (2\pi)^4
\delta^4 (p_1 + p_2 - p_3 - p_4 - k_S) E_S f_1^{(e)} f_2^{(N_i)} P_{\rm decay} P_{\rm abs}  \,, \nonumber \\ &&
\end{eqnarray}
where $\mathcal{M}_i$'s are the coherent scattering amplitudes for the nuclei $N_i$,
and $f_{1,\,2}$ are the non-relativistic Maxwell-Boltzmann distributions of the incoming electron and nucleons in the non-degenerate limit in the Sun (cf.~(\ref{eqn:f12})).
Following the calculations in Ref.~\cite{Dev:2020jkh}, the emission rate in Eq.~(\ref{eqn:rate:master})  can be simplified to be
\begin{eqnarray}
\label{eqn:QB:eN}
{\cal Q}_{}^{} (r,\phi) &=& \Big( \sum_i Z_{N_i}^2 A_{N_i}^2 n_{N_i} (r) \Big)
\frac{ \alpha^2 y_{N}^2 \sin^2\theta T^{1/2}(r) n_e (r) }{\pi^{3/2} m_e^{3/2}}  \nonumber \\
&& \times \int_{q}^{\infty} {\rm d}u \int_0^\infty {\rm d}v \int_q^\infty {\rm d}x \int_{-1}^1 {\rm d}z  \sqrt{uv} e^{-u} \sqrt{x^2 - q^2} \frac{\delta (u-v-x)}{(u+v-2\sqrt{uv}z)^2} P_{\rm decay} P_{\rm abs}  \,, \nonumber \\ &&
\end{eqnarray}
where $\alpha={e^2}/{4\pi}$ is the fine-structure constant, and we have summed up the coherent contributions from all the nuclei elements $N_i$, with $Z_{N_i}$, $A_{N_i}$ and $n_{N_i}(r)$ the corresponding atomic number, mass number and number density in the Sun, respectively. $n_{N_i}(r)$ is related to the density $\rho(r)$ and mass fraction $Y_{N_i}(r)$ in the Sun via
\begin{eqnarray}
n_{N_i}(r) \simeq \frac{\rho(r) Y_{N_i} (r)}{A_{N_i} m_N} \,.
\end{eqnarray}
$n_e(r)$ in Eq.~(\ref{eqn:QB:eN}) is the number density of electrons in the Sun, and can be easily obtained from the radial nuclei density $n_{N_i}(r)$ in conjunction with imposing local electric charge neutrality
\begin{eqnarray}
n_e (r) = \sum_i Z_{N_i} n_{N_i} (r) \,.
\end{eqnarray}

\begin{figure}[!t]
    \centering
    \includegraphics[width=0.48\textwidth]{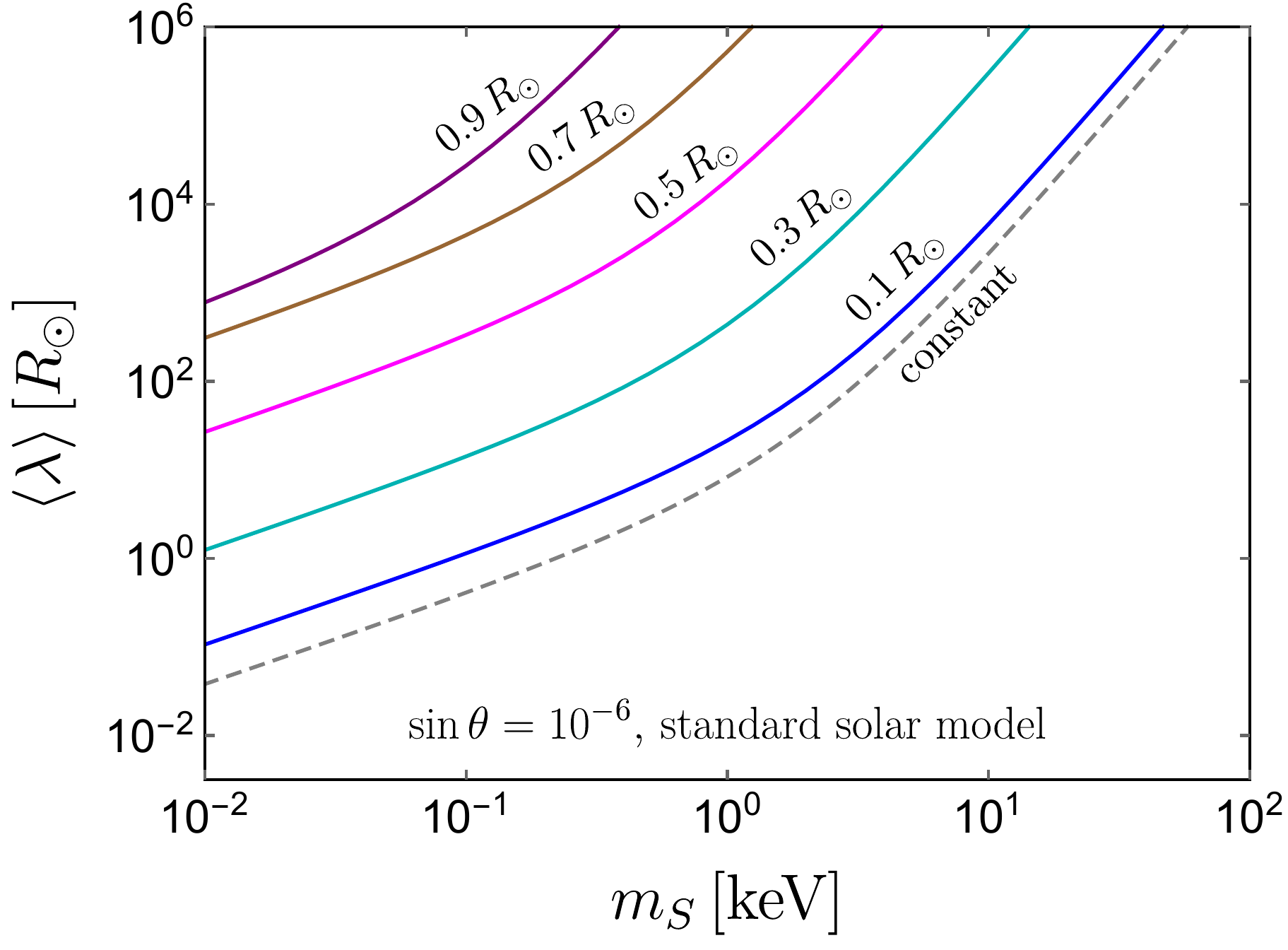}
    \caption{The MFP $\langle \lambda_{} \rangle$ of $S$ in the Sun, as function of its mass $m_S$ at the radial positions $r=[0.1,\, 0.3,\, 0.5,\, 0.7,\, 0.9]R_\odot$ within the standard solar model. The dashed line is the case with constant solar temperature $T= 1$ keV and constant electron density $n_{e} = 10^{26} \, {\rm cm}^{-3}$.  The scalar mixing is fixed to $\sin\theta=10^{-6}$. }
    \label{fig:solarmfps}
\end{figure}

For the decay factor $P_{\rm decay}$ in Eq.~(\ref{eqn:rate:master}), we follow \eqref{eqn:Rdecay} and the decay width here is defined as $\Gamma_S = (m_S/E_S) \Gamma_0 (S\to\gamma\gamma)$, with the proper decay width
\begin{eqnarray}
\label{eqn:width}
\Gamma_0 (S \to \gamma\gamma) = \frac{121}{9}  \frac{\alpha^2 m_S^3 \sin^2\theta}{512 \pi^3 v_{\rm EW}^2} \,,
\end{eqnarray}
with the factor of ${121}/{9}$ coming from summing up the loop factors for all the charged SM particles in the limit of $m_S \to 0$.
The expression for the absorption factor $P_{\rm abs}$ in Eq.~(\ref{eqn:rate:master}) is the same as Eq.~(\ref{eqn:Rabs}),
and the most important absorption process of $S$ is the inverse $e-N_i$ bremsstrahlung channel $e + N_i + S \to e + N_i$. The corresponding formula  is similar to  Eq.~(\ref{eqn:mfp}), and turns out to be~\cite{Dev:2020jkh}
\begin{eqnarray}
\label{eqn:MFPB}
\lambda^{-1}_{} (r;x) & = & \left( \sum_i Z_{N_i}^2 A_{N_i}^2 n_{N_i} (r) \right) \frac{1}{x^2} \frac{2 \pi^{1/2}  \alpha^2 y_{N}^2 \sin^2\theta n_e(r)}{ m_e^{3/2} T^{7/2}(r)}  \nonumber \\
&& \times \int_{0}^{\infty} {\rm d}u \int_q^\infty {\rm d}v \int_{-1}^1 {\rm d}z \sqrt{uv} e^{-u} \frac{\delta (u-v+x)}{(u+v-2\sqrt{uv}z)^2} \,,
\end{eqnarray}
where we have summed over all the nuclei for coherent scattering. As in Eq.~(\ref{eqn:lambda:average}), averaging over the energy dependence for the MFP in Eq.~(\ref{eqn:MFPB}), we can get the energy-independent MFP $\langle \lambda_{}^{} \rangle (r)$, which is also a function of $r$. For illustrative purposes, the dependence of $\langle \lambda_{}^{} \rangle$ on the scalar mass $m_S$ is shown in Fig.~\ref{fig:solarmfps}. Five benchmark radial positions $r=[0.1,\, 0.3,\, 0.5,\, 0.7,\, 0.9]R_\odot$ from the core of the Sun have been chosen, and the scalar mixing has once again been fixed to $\sin\theta=10^{-6}$. As a comparison, the MFP with the constant solar temperature $T = 1$ keV and electron density $n_e = 10^{26} \, {\rm cm}^{-3}$ is shown in Fig.~\ref{fig:solarmfps} by the dashed gray line, assuming a mass fraction of 75\% for Hydrogen and 25\% for Helium-4. It is clear that the MFP with the standard solar model can be a few times larger than the constant profile case in the solar center. In the outer layers, the MFP grows rapidly as the temperature and density drop significantly.

\subsection{Results}
\label{sec:sun:results}

\begin{figure}[!t]
    \centering
     \includegraphics[width=0.6\textwidth]{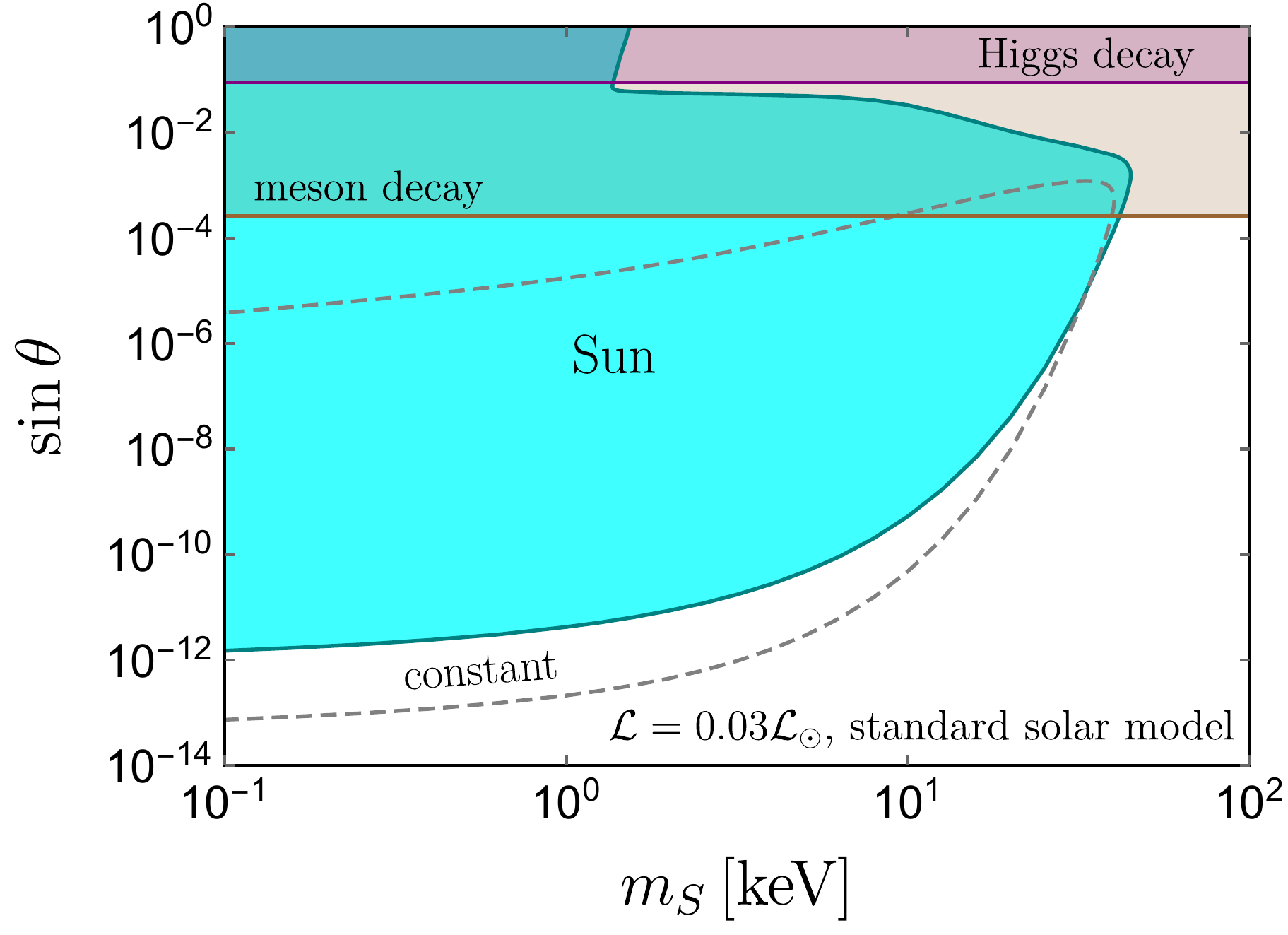}
    \caption{Solar bounds on the scalar mass $m_S$ and mixing angle $\sin\theta$, using 3\% of the measured solar luminosity $\mathcal{L}_\odot$ (cyan). The dashed gray line indicates the solar limit with constant solar temperature $T= 1$ keV and constant electron density $n_{e} = 10^{26} \, {\rm cm}^{-3}$~\cite{Dev:2020jkh}. Also shown are the constraints from invisible meson decays (gray)~\cite{Dev:2017dui, Egana-Ugrinovic:2019wzj, Dev:2019hho} and the current LHC limit on invisible decay of the SM Higgs boson (purple) \cite{CMS:2018yfx,ATLAS-CONF-2020-008}.  }
    \label{fig:sunlimits}
\end{figure}

The luminosity due to emission of the scalar $S$ from the Sun is
\begin{align}
{\cal L}_S = \sum_i \int {\rm d}V_\odot {\cal Q}_i  \,.
\label{eqn:lumi}
\end{align}
As in the supernova case in Section~\ref{sec:supernova}, we consider the standard solar model and the geometry in the decay and absorption of $S$.
To set exclusion bounds on the scalar mass $m_S$ and mixing angle $\sin\theta$, we adopt 3\% of measured solar luminosity, which is from the combination of helioseismology (speed of sound, surface helium and convective radius) and solar neutrino observations~\cite{Song:2017kvf, Guarini:2020hps}
\begin{eqnarray}
{\cal L}_\odot^{\rm (limit)} = 1.2 \times 10^{32} \ {\rm erg/sec} \,.
\end{eqnarray}
The resultant excluded region of $m_S$ and $\sin\theta$ is shown in Fig.~\ref{fig:sunlimits} as the cyan shaded region. The solar limit with constant temperature $T=1$ keV and electron density $n_e = 10^{26} \, {\rm cm}^{-3}$ is presented as the dashed gray line. Comparing the solar limits with standard solar model and the constant profile case, the limits on the mixing angle is moved upward.
This is due to the following reasons:
\begin{itemize}
    \item In most regions of the Sun, the temperature is below 1 keV (cf. the upper left panel of Fig.~\ref{fig:solarprofies}), and the electron number density is below the value of $10^{26} \, {\rm cm}^{-3}$ (cf. the right panel of Fig.~\ref{fig:solarprofies}). As a result, less $S$ is produced when the standard solar model is adopted, compared to the constant profile case. This leads to the exclusion of $\sin\theta$ down to $1.5\times10^{-12}$ (instead of $7.4\times10^{-14}$) when the scalar is light.
    \item As shown in Fig.~\ref{fig:solarmfps}, the MFP of $S$ is significantly longer in the standard solar model than in the constant profile case. In addition, the energy dependence of $\lambda (r;x)$ also plays an important role. In total, the absorption of $S$ is actually much weaker than expected in the constant profile case. Therefore, a larger mixing angle $\sin\theta$ can be excluded even up to order one, compared to $1.2\times10^{-3}$ in the constant profile case.
    \item With the standard solar model, the scalar mass can be excluded up to 45 keV, which is close to the value of 40 keV in the constant profile case.
\end{itemize}
It should be noted  that the MFP $\lambda(r;x)$ in Eq.~(\ref{eqn:MFPB}) does not depend on the scalar mass $m_S$, which is very different from the supernova case (cf. Eq.~(\ref{eqn:mfp2})). However, the MFP in the Sun depends on the scalar energy $E_S>m_S$; when we integrate over $x = E_S/T$ to calculate the emission rate in Eq.~(\ref{eqn:QB:eN}), the effect of scalar mass on the integration range of $x > q = m_S/T$ is important. This explains the nontrivial feature in the upper boundary of the excluded cyan region in Fig.~\ref{fig:sunlimits}.

As demonstrated in the right panel of Fig.~\ref{fig:snlimits}, the mixing angle $\sin\theta$ is stringently constrained by high-precision collider data, cosmological observations and astrophysical constraints. In particular, even if in the limit of $m_S \to 0$, the FCNC limits on $S$ do not disappear, but instead approach a constant value of $\simeq 3\times 10^{-4}$~\cite{Dev:2017dui, Egana-Ugrinovic:2019wzj, Dev:2019hho}. This is indicated by the gray shaded region in Fig.~\ref{fig:sunlimits}. At the LHC, the mixing angle is also constrained by the invisible decay of the SM Higgs $h$~\cite{CMS:2018yfx,ATLAS-CONF-2020-008}, which excludes $\sin\theta > 0.09$~\cite{Dev:2020jkh}. This is shown as the purple shaded region in Fig.~\ref{fig:sunlimits}. The cosmological limits on $S$ depend on the reheating temperature $T_R$. It is expected that, within standard cosmology, the whole parameter space in Fig.~\ref{fig:sunlimits} has been excluded by the cosmological constraints~\cite{Berger:2016vxi, Fradette:2018hhl, Ibe:2021fed}. However, the cosmological limit on $S$ might be dramatically altered for a non-standard cosmological history in the early Universe~\cite{Bernal:2018kcw}. In such a case, the solar limit on $S$, as well as those from WDs and RGs in the upcoming section, will provide solid, independent limits on $S$ in the low mass range.

\section{Updated limits from white dwarfs and red giants}
\label{sec:others}

The limits derived from SN1987A and the Sun with radial stellar profiles computed from the equation-of-state can also be applied to other stars, notably WDs and RGs. However, the profiles of these stars suffer from much larger uncertainties. Therefore, we will assume the baryon densities in these stars are constants as in Ref.~\cite{Dev:2020jkh}. However, we will consider the dependence of the decay and absorption factors on the geometric parameters $r$ and $\phi$ as in Section~\ref{sec:sn:calculation}. The dominant elements in the stellar cores and their mass fractions, temperatures $T$, electron number densities $n_e$ and sizes $R$ are displayed for convenience in Table~\ref{tab:stars}~\cite{Dev:2020jkh}.

\begin{table}[!t]
    \centering
      \caption{Stellar parameters for the RGs and WDs~\cite{Dev:2020jkh}: the dominant elements in the core and their mass fractions, core temperatures $T$, electron number densities $n_e$, sizes $R$, and the luminosity limits in unit of the solar luminosity of ${\cal L}_\odot = 4 \times 10^{33}$ erg/sec. }
    \label{tab:stars}
    \vspace{0.2cm}
    \begin{tabular}{l|c|c|c|c|c}
    \hline\hline
    Star & Core composition & $T$ [keV]  & $n_e \; [{\rm cm}^{-3}]$  & $R$ [cm] & ${\cal L}/{\cal L}_\odot$ \\ \hline
    RGs~\cite{Davidson:1991si, Viaux:2013lha}  & $^{4}$He  & 10 & $3\times 10^{27}$  & $3\times 10^{9}$ &
    2.8 \\ \hline
    \multirow{2}{*}{WDs~\cite{Raffelt:1996wa, Blinnikov:1990ui, Panotopoulos:2020kuo, whitedwarf}}   & 50\% $^{12}$C   &
    \multirow{2}{*}{6}  & \multirow{2}{*}{$10^{30}$}   &
    \multirow{2}{*}{$10^{9}$}  & \multirow{2}{*}{$10^{-5}$ } 
    \\ 
    & 50\% $^{16}$O &&& \\ \hline\hline
    \end{tabular}
\end{table}

\begin{figure}[!t]
    \centering
    \includegraphics[width=0.6\textwidth]{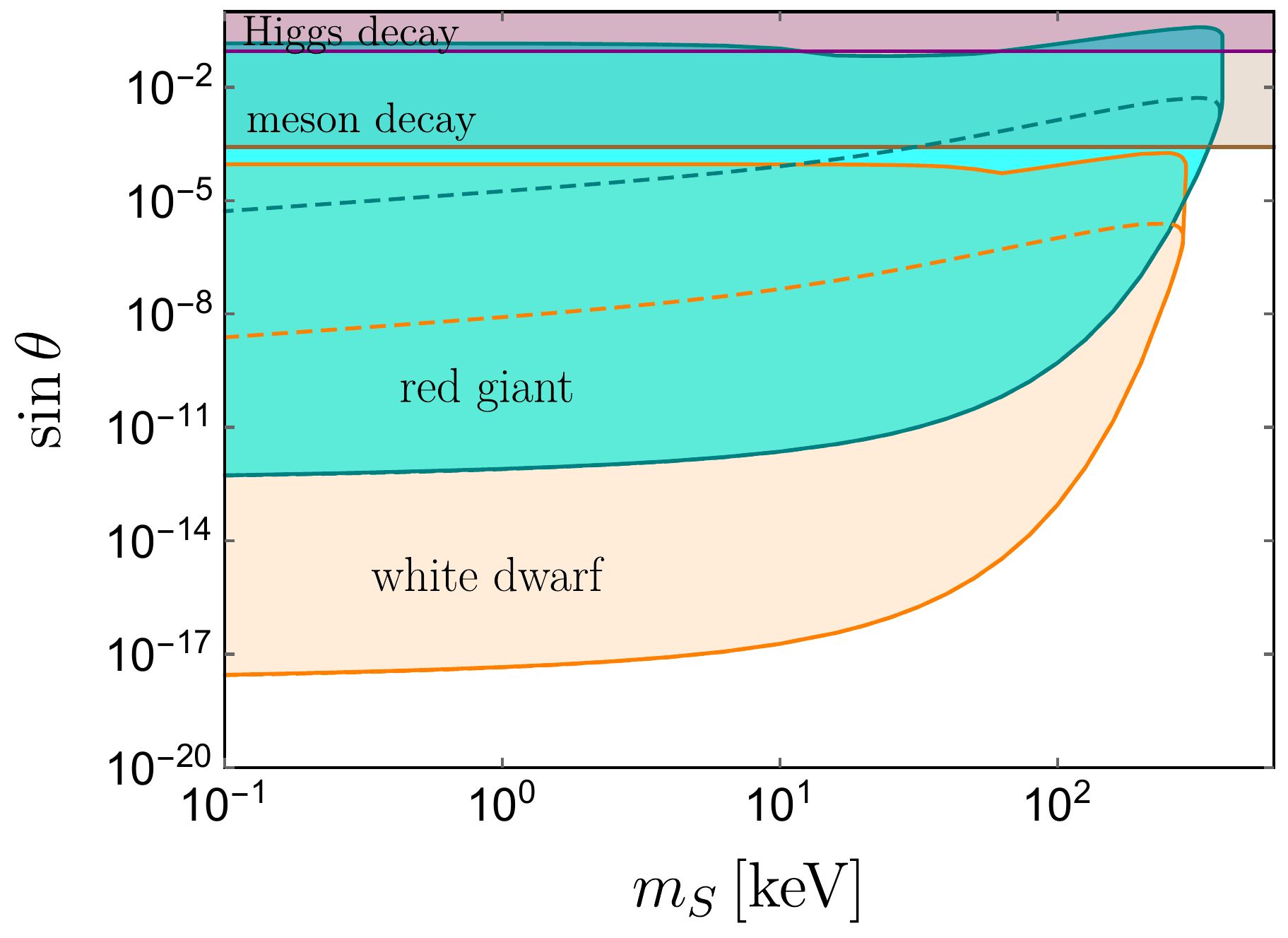}
    \caption{
    Luminosity bounds on the scalar mass $m_S$ and mixing angle $\sin\theta$ from WDs (cyan) and RGs (orange) with constant density, temperature and mass fractions given in Table~\ref{tab:stars}. The dashed cyan and orange lines are, respectively, the corresponding limits for WDs and RGs from Ref.~\cite{Dev:2020jkh} without the geometry taken into consideration. Also shown are the constraints from invisible meson decays (gray)~\cite{Dev:2017dui, Egana-Ugrinovic:2019wzj, Dev:2019hho} and the current LHC limit on invisible decay of the SM Higgs boson (purple) \cite{CMS:2018yfx,ATLAS-CONF-2020-008}.  }
    \label{fig:starlimits}
\end{figure}

The calculation procedure of the stellar limits is quite similar to the case of the Sun in Section~\ref{sec:Sun}, except the profiles of densities, temperatures and mass fractions. Applying the stellar limits in the last column of Table~\ref{tab:stars}, the resultant bounds on $m_S$ and $\sin\theta$ from these stars are presented in Fig.~\ref{fig:starlimits}. The WD and RG limits are shown as the cyan and purple shaded regions, respectively. The scalar mass is excluded up to 392 keV and 290 keV by RGs and WDs, respectively, improving slightly the limits of 384 keV and 283 keV from Ref.~\cite{Dev:2020jkh}. The luminosity constraints from WDs are very stringent, down to $10^{-5} \; {\cal L}_\odot$, which enable us to exclude a significantly large parameter space of $\sin\theta$, ranging from  $2.8\times10^{-18}$ up to $1.8\times10^{-4}$, compared to the range of $2.8\times10^{-18}$ to $2.4\times10^{-6}$ without the geometry taken into account. The luminosity constraint from RGs is relatively weak, excluding the mixing angle from $5.3 \times10^{-13}$ to $0.39$, with the range of $5.3\times10^{-13}$ to $5.3\times10^{-3}$ without the geometry.
The lower bounds of the excluded regions are mainly determined by the emission rate ${\cal Q}$, which are hardly affected by inclusion of the geometric factor. The upper bounds of the excluded regions are dominated by the absorption of $S$ in the stars. When the geometry is included, a larger mixing angle can be excluded.
Compared to RGs, a smaller luminosity is excluded by WDs, therefore at small mixing angles the WDs outperform the RGs significantly. However, when the mixing angle is large, the absorption of $S$ in the RGs are weaker than in the WDs, thus the RGs can exclude a larger mixing angle.
As in Fig.~\ref{fig:sunlimits} for the solar case, the limits from WDs and RGs are also complementary to those from collider data, cosmology and other astrophysical limits. The limits from meson data and invisible Higgs decay are also shown in Fig.~\ref{fig:starlimits} as the shaded gray and purple regions, respectively.

\section{Conclusion}
\label{sec:conclusion}

\begin{table}[!t]
    \centering
      \caption{Comparison of the limits on CP-even scalar $S$ from SN1987A, the Sun, RGs and WDs obtained in this paper including the geometric factor and stellar profiles to those with constant temperature, density and mass fractions~\cite{Dev:2020eam, Dev:2020jkh}. The second and third columns indicate the stellar profile adopted and whether geometry is included, respectively. The fourth and fifth columns are the excluded ranges of $\sin\theta$ and $m_S$, respectively. See text and Figs.~\ref{fig:snlimits}, \ref{fig:sunlimits} and \ref{fig:starlimits} for more details.}
    \label{tab:comparison}
    \vspace{0.2cm}
    \begin{tabular}{c|c|c|c|c} \hline\hline
    Star & Profile & Geometry  & $\sin\theta$ range  & $m_S$ range \\ \hline
    \multirow{4}{*}{SN1987A}  & $-$ & $-$ & $2.4\times10^{-7} - 9.0\times10^{-6}$ & $<249$ MeV \\
    & Fischer 11.8$M_\odot$  & \checkmark & $1.5\times 10^{-7} - 3.8\times10^{-5}$  & $<187$ MeV \\
    & Fischer 18$M_\odot$  & \checkmark & $1.3\times 10^{-7} - 3.1\times10^{-5}$  & $<219$ MeV \\
    & Nakazato 13$M_\odot$  & \checkmark & $1.5\times 10^{-7} - 3.6\times10^{-5}$  & $<205$ MeV \\ \hline
    \multirow{2}{*}{Sun}   & $-$  & $-$  & $7.4\times10^{-14} - 1.2\times10^{-3}$  & $<40$ keV \\
    & standard solar model & \checkmark & $1.5\times10^{-12} - 1$ & $<45$ keV \\ \hline
    \multirow{2}{*}{RGs}   & $-$  & $-$  & $5.3\times10^{-13} - 5.3\times10^{-3}$ & $<384$ keV \\
    & $-$ & \checkmark & $5.3\times10^{-13} - 0.39$ & $<392$ keV \\ \hline
    \multirow{2}{*}{WDs}   & $-$  & $-$ & $2.8\times10^{-18} - 2.4\times10^{-6}$ & $<283$ keV \\
    & $-$ & \checkmark & $2.8\times10^{-18} - 1.8\times10^{-4}$ & $<290$ keV \\ \hline
    \hline
    \end{tabular}
\end{table}

Compact astrophysical objects such as the supernova cores and stars such as the Sun, RGs and WDs offer some of the densest environments in the Universe. Therefore, they provide some of the primary high-intensity facilities in nature to search for the feeble interactions of light BSM particles, such as ALPs and light CP-even scalars. Here we have considered a generic CP-even scalar $S$ mixing with the SM Higgs, so that we have only two free parameters, i.e. the scalar mass $m_S$ and the mixing angle $\sin\theta$. We have updated the existing stellar limits~\cite{Dev:2020eam, Dev:2020jkh} on the $S$ mass and mixing from SN1987A, the Sun, RGs and WDs by including the stellar profiles (for the supernova and the Sun) and the geometric effects for the decay and absorption of $S$ (cf. Fig.~\ref{fig:geometry}). For SN1987A, we adopt  three benchmark profiles, i.e. Fischer 11$M_\odot$, Fischer 18$M_\odot$ and Nakazato 13$M_\odot$ (see Fig.~\ref{fig:sn1987aradialprofiles}), while for the Sun we take the standard solar model (see Fig.~\ref{fig:solarprofies}). The comparison of the excluded ranges of $\sin\theta$ and $m_S$ are collected in Table~\ref{tab:comparison}. It is clear from this table that a proper implementation of the stellar profiles and geometry is essential to extract precise astrophysical limits on $S$.

As the densities in the centers of SN1987A and the Sun are higher than at the surfaces, the MFPs of $S$ in the centers are shorter, as shown in Figs.~\ref{fig:snmfp} and \ref{fig:solarmfps}. Considering both decay and absorption of $S$ in the supernova core, and requiring the luminosity due to the emission of $S$ is smaller than the inferred SN1987A luminosity of $3\times10^{52}$ erg/sec, the supernova limits with the three different profiles are found to be similar, excluding the mixing angle in the range of $\simeq 1\times 10^{-7}$ to $\sim 3\times 10^{-5}$, which is nonetheless significantly broader than the range of $2.4\times10^{-7}$ to $9.0\times10^{-6}$ in the constant profile case~\cite{Dev:2020eam}. As shown in Figure.~\ref{fig:snlimits}, the scalar mass can be excluded up to 219 MeV, which is close to the constant profile case. In spite of the profile uncertainties, the supernova limits are largely complementary to those from colliders, cosmology and other astrophysical limits. In particular, the supernova limits have totally excluded the blind spot left by the cosmological limit with a high reheating temperature $T_R \gtrsim 100$ GeV~\cite{Ibe:2021fed}.

For the Sun, using 3\% of the observed solar luminosity to set limits on $S$, the solar limit on the mixing is updated from the range of $7.4\times10^{-14}$ to $1.2\times10^{-3}$ upward to the range of $1.5\times10^{-12}$ up to 1. The scalar mass is excluded up to 45 keV, close to the constant profile case~\cite{Dev:2020jkh} (see Figure.~\ref{fig:sunlimits}).

As for RGs and WDs, the current uncertainties of their profiles are very large, and therefore, we assume their densities, temperatures and element mass fractions are constants, and consider only the geometric effects. The updated limits exclude the mixing angle in the range of $5.3\times10^{-13}$ to $0.39$ for RGs and $2.8\times10^{-18}$ to $1.8\times10^{-4}$ for WDs, respectively, with scalar mass up to, respectively, 392 keV and $290$ keV (see Fig.~\ref{fig:starlimits}), broadening significantly the limits in the constant profile cases~\cite{Dev:2020jkh}.

The procedure for calculating stellar limits on the light scalar $S$ in this paper can be applied to other astrophysical systems such as neutron star mergers~\cite{Dev:2021kje} and brown dwarfs~\cite{Nakajima:1995sv}, as well as to other BSM particles such as the dark photon, $Z'$ boson, ALPs and dark particles. In the era of multi-messenger astronomy and precision cosmology, it is essential to perform more precise theoretical calculations of these astrophysical limits such as the ones reported here.
In this paper, we have limited ourselves only to the luminosity considerations to derive the limits. However, depending on the mass and coupling of the BSM particles, their decay outside the stars could generate MeV-scale $\gamma$-rays~\cite{VonFeilitzsch:1988ip,Kolb:1988pe,Dodelson:1992tv,Bludman:1992ww,Dodelson:1993ms,Oberauer:1993yr,Brockway:1996yr,Berezhiani:1999qh,Gianfagna:2004je,Kazanas:2014mca,Payez:2014xsa,DeRocco:2019njg,Fiorillo:2021gsw,Caputo:2021rux}, which will provide additional limits on the BSM particles. Similarly, if $S$ decays into photons outside the Sun, it will contribute to the X-ray flux at the scale of ${\cal O}$(keV) to ${\cal O}$(10 keV) from the solar flares and flare-like brightenings, and thus could get constrained by the NuSTAR data~\cite{Grefenstette:2016esh, Marsh:2017ait, Kuhar:2018lao}. In the future, the solar X-ray limits on $S$ can be further improved by the SSAXI observations~\cite{Hong:2018yrt}. These multi-messenger limits will be reported in a future publication.

\section*{Acknowledgments}

We thank Rabi Mohapatra for helpful discussions and collaboration on earlier related works. We are also grateful to Shmuel Nussinov for enlightening discussions. S.B. is supported by funding from the European Union¡¯s Horizon 2020 research and innovation programme under grant agreement No 101002846 (ERC CoG ``CosmoChart'') as well as support from the Initiative Physique des Infinis (IPI), a research training program of the Idex SUPER at Sorbonne
Universit\'{e}. The work of B.D. is supported in part by the U.S. Department of Energy under grant No.~DE-SC0017987 and by a Fermilab Intensity Frontier Fellowship. Y.Z. is supported by the National Natural Science Foundation of China under grant No.\  12175039, the 2021 Jiangsu Shuangchuang (Mass Innovation and Entrepreneurship) Talent Program No.\ JSSCBS20210144, and the ``Fundamental Research Funds for the Central Universities.''

\bigskip
\noindent
{\bf Note added}: While this paper was being finalized,  Ref.~\cite{Caputo:2022rca} appeared on arXiv, where they also considered the geometric effect for the decay and absorption of axion-like particles inside the stellar core. One may notice the similarity between Fig.~\blue{7} in Ref.~\cite{Caputo:2022rca} and Fig.~\ref{fig:geometry} in this paper, and similar dependence of MFP on the geometry, but our results were derived independently before Ref.~\cite{Caputo:2022rca} appeared.

\appendix
\section{Additional functions}
\label{sec:appendix}

Following Refs.~\cite{Giannotti:2005tn, Dent:2012mx}, the dimensionless functions are defined to be:
\begin{align}
\label{eqn:uvxyqz}
& u \ \equiv \ \frac{{\bf p}_i^2}{m_N T} \,, \qquad
 v \ \equiv \ \frac{{\bf p}_f^2}{m_N T} \,, \qquad
 x \ \equiv \ \frac{E_S}{T} \,,  \qquad
 q \ \equiv \ \frac{m_S}{T} \,, \qquad
 y \ \equiv \ \frac{m_{\pi}^2}{m_N T} \,.
\end{align}
Denoting ${\bf p}_i$ (with $i=1,\, 2,\, 3,\, 4$) as the three-momenta for the two nucleons in the initial state of the nucleon bremsstrahlung process and the two nucleons in the final state, we define
\begin{align}
& \quad {\bf p}_1  \ \equiv \  {\bf P} + {\bf p}_i \,, \quad
{\bf p}_2 \ \equiv \ {\bf P} - {\bf p}_i \,, \quad 
{\bf p}_3  \ \equiv \  {\bf P} + {\bf p}_f \,, \quad
{\bf p}_4 \ \equiv \ {\bf P} - {\bf p}_f \,,
\end{align}
where ${\bf P}$ is the total three-momentum in the initial and final states, ${\bf p}_{i,f}$ are the relative three-momenta between the nucleons in the initial and final states, respectively, and $z\equiv \cos\theta_{if}$ is defined as the cosine of the angle between ${\bf p}_i$ and ${\bf p}_f$. The dimensionless function ${\cal I}_{\rm tot}$ in Eq.~(\ref{eqn:rate2}) is
\begin{eqnarray}
\label{eqn:Itot}
{\cal I}_{\rm tot} \ = \
\frac14 y_{hNN}^2 \left( \frac{q}{x} \right)^4 {\cal I}_A +
\frac{1}{81} \left( \frac{m_N}{v_{\rm EW}} \right)^2 {\cal I}_B +
\frac{1}{9} y_{hNN} \left( \frac{q}{x} \right)^2
\left( \frac{m_N}{v_{\rm EW}} \right) {\cal I}_C \,.
\end{eqnarray}
Note that this expression differs from that in Ref.~\cite{Dev:2020eam} by a factor of 1/4 and 1/2 for the first and third terms, respectively, which is due to the higher-order corrections in expansion of the small parameters~\cite{Dev:2021kje}. Decomposing further into the $pp$, $nn$ and $np$ contributions, we obtain
\begin{eqnarray}
\label{eqn:Iabc}
{\cal I}_{A,\, B,\, C} \ = \ {\cal I}_{A,\, B,\, C}^{(pp)} + {\cal I}_{A,\, B,\, C}^{(nn)} + 4{\cal I}_{A,\, B,\, C}^{(np)} \,,
\end{eqnarray}
with ${\cal I}_{A,\, B,\, C}^{(pp)} = {\cal I}_{A,\, B,\, C}^{(nn)}$, and
\begin{eqnarray}
\label{eqn:IApp}
{\cal I}_{A}^{(pp)} & \ = \ &
\frac{c_k^2}{c_{k\pi}^2} + \frac{c_l^2}{c_{l\pi}^2} - \frac{c_{kl}^2}{c_{k\pi}c_{l\pi}} \,, \\
\label{eqn:IBpp}
{\cal I}_{B}^{(pp)} (r) & \ = \ &
\left( q^2 \frac{T(r)}{m_N} + \frac{11}{2} y \right)^2
\left[ \frac{c_k^2}{c_{k\pi}^4} + \frac{c_l^2}{c_{l\pi}^4} - \frac{c_{kl}^2}{c_{k\pi}^2c_{l\pi}^2} \right] \,, \\
\label{eqn:ICpp}
{\cal I}_{C}^{(pp)} (r) & \ = \ &
\left( q^2 \frac{T(r)}{m_N} + \frac{11}{2} y \right)
\left[ \frac{c_k^2}{c_{k\pi}^3} + \frac{c_l^2}{c_{l\pi}^3} - \frac{c_{kl}^2}{2c_{k\pi}c_{l\pi}^2} - \frac{c_{kl}^2}{2c_{k\pi}^2c_{l\pi}} \right] \,, \\
\label{eqn:IAnp}
{\cal I}_{A}^{(np)} & \ = \ &
\frac{c_k^2}{c_{k\pi}^2} + \frac{4c_l^2}{c_{l\pi}^2} + \frac{2c_{kl}^2}{c_{k\pi}c_{l\pi}} \,, \\
\label{eqn:IBnp}
{\cal I}_{B}^{(np)} (r) & \ = \ &
\left( q^2 \frac{T(r)}{m_N} + \frac{11}{2} y \right)^2
\left[ \frac{c_k^2}{c_{k\pi}^4} + \frac{4c_l^2}{c_{l\pi}^4} + \frac{2c_{kl}^2}{c_{k\pi}^2c_{l\pi}^2} \right] \,, \\
\label{eqn:ICnp}
{\cal I}_{C}^{(np)} (r) & \ = \ &
\left( q^2 \frac{T(r)}{m_N} + \frac{11}{2} y \right)
\left[ \frac{c_k^2}{c_{k\pi}^3} + \frac{4c_l^2}{c_{l\pi}^3} + \frac{c_{kl}^2}{c_{k\pi}c_{l\pi}^2} + \frac{c_{kl}^2}{c_{k\pi}^2c_{l\pi}} \right] \,.
\end{eqnarray}
The $c$ functions are defined as:
\begin{eqnarray}
c_k & \ \equiv \ &  u+v-2z\sqrt{uv}  \,, \\
c_{k\pi} & \ \equiv \ &  u+v+y-2z\sqrt{uv} \,, \\
c_l & \ \equiv \ &  u+v+2z\sqrt{uv} \,, \\
c_{l\pi} & \ \equiv \ &  u+v+y+2z\sqrt{uv} \,, \\
c_{kl}^2 & \ \equiv \ &
 u^2 + v^2 + 2 u v (-3 + 2 z^2) \,.
\end{eqnarray}

\section{Subdominant production channels in the Sun}
\label{sec:appendix2}

For the production of $S$ via the Compton-like process $e+\gamma \to e + S$, the total cross section is given by~\cite{Araki:1950}
\begin{eqnarray}
\label{eqn:Compton:total}
\sigma_{\rm C} &=& \frac{\alpha y_e^2 \sin^2\theta}{3m_e^2}
f_{\rm C}(q',y') \, .
\end{eqnarray}
where
the dimensionless parameters $q' \equiv m_S/m_e$,
$y' \equiv E_\gamma/m_e$ (with $E_\gamma$ being the photon energy), and
the dimensionless function $f_{\rm C} (q,y)$ is defined as~\cite{Dev:2020jkh}
\begin{eqnarray}
\label{eqn:fC}
f_{\rm C} (q,y) &\equiv& \frac{3}{16 y^3 (1+2y)^2}
\bigg\{ 2\tilde{y} \Big[ -2 (2+3y) (2+5y+y^2) + q^2 (2+8y+7y^2) \Big]  \nonumber \\
&& +(1+2y)^2  \Big( 2 (2+y)^2 -2q^2 (3+y) + q^4 \Big) \log \left( \frac{2(1+y +\tilde{y}) - q^2 }{2(1+y -\tilde{y}) - q^2 } \right) \bigg\} \,,
\end{eqnarray}
with
\begin{eqnarray}
\tilde{y} \equiv \sqrt{\left[ y + q \left( 1-\frac12 q \right) \right] \left[ y - q \left( 1+\frac12 q \right) \right] } \,.
\label{eq:ytilde}
\end{eqnarray}
In the limit of $y' \to 0$ and $q' \to 0$, $f_{\rm C}(0,0)=1$. As the electron mass $m_e \gg T \sim$ keV,
for simplicity we have neglected the kinetic energy of electrons and assumed the electrons to be at rest in the initial state. With the approximation $E_\gamma \simeq E_S$, the energy emission rate in the Compton-like process is
\begin{eqnarray}
\label{eqn:rate:compton}
{\cal Q}_{\rm C} (r) \simeq n_e (r) \int \frac{2 d^3 {\bf k}_\gamma}{(2\pi)^3} \frac{E_\gamma}{e^{E_\gamma/T(r)}-1}
\sigma_{\rm C} (q',y')  \,,
\end{eqnarray}
where ${\bf k}_\gamma$ is the three-momentum of photon.  

For the Primakoff process $\gamma + X \to X + S$, the coherent production cross section is~\cite{Dicus:1978fp}
\begin{eqnarray}
\label{eqn:xs:Primakoff}
\sigma_{{\rm P},\, X} (r) =  64\pi Z_X^2 \alpha  \frac{E_\gamma \Gamma_0 (S\to \gamma\gamma)}{m_S^2}
\frac{\sqrt{E_\gamma^2 - m_S^2} (E_\gamma - m_S)}{(m_S^2 + 2 m_S E_\gamma + k_{\rm scr}^2(r))^2} \,,
\end{eqnarray}
where $Z_X$ is the atomic number of the nucleus $X$ (for electrons $Z_X=1$), and the scalar decay width $\Gamma_0 (S \to \gamma\gamma)$ is given in Eq.~(\ref{eqn:width}). To remove the divergence in Eq.~(\ref{eqn:xs:Primakoff}) in the limit of massless $S$, the screening scale $k_{\rm scr} (r)$ is introduced to the propagator~\cite{Raffelt:1996wa}:
\begin{eqnarray}
k^2_{\rm scr}(r) = \frac{4\pi\alpha}{T} \frac{\rho(r)}{m_N} \left(Y_e(r) + \sum_j Z_{j}^2 Y_{j}(r) \right) \,,
\end{eqnarray}
where
$Y_{e,\,j}(r)$ are the number fractions of electrons and the baryons $j$. Then the energy loss rate for the Primakoff process is
\begin{eqnarray}
\label{eqn:rate:Primakoff}
{\cal Q}_{\rm P} (r) = \sum_X n_X(r) \int \frac{2 d^3 {\bf k}_\gamma}{(2\pi)^3} \frac{E_S}{e^{E_\gamma/T(r)}-1}  \sigma_{{\rm P},\, X}(r) \,.
\end{eqnarray}
where we have summed up the contributions from all the incident electrons or nucleus.

The resonant plasma contribution is dominated by the coupling of $S$ to electrons, and will be efficient when the scalar mass $m_S$ is smaller than the plasma frequency $\omega_p$, i.e.
$m_S<\omega_p<T$. The corresponding production rate is~\cite{Hardy:2016kme}
\begin{eqnarray}
\label{eqn:Qplm}
{\cal Q}_{\rm pl}(r) \ \simeq \ \frac{y_e^2\sin^2\theta} {16\pi^2\alpha} k^2_{\omega_p}(r) \omega_p^3(r)
\frac{1}{e^{\omega_p(r)/T(r)}-1} \,,
\end{eqnarray}
where the plasma frequency $\omega_p(r) \simeq \sqrt{e^2 n_e(r)/m_e}$, and $k^2_{\omega_p}(r) = \omega_p^2(r) - m_S^2$.

\bibliographystyle{JHEP}
\bibliography{refs}
\end{document}